\documentclass[%
amsmath,amssymb,
aps,
pre,
]{revtex4-1}
\usepackage{color}
\usepackage{graphicx}
\usepackage{dcolumn}
\usepackage{bm}
\usepackage{subfigure}

\usepackage{ulem}

\begin{document}


\preprint{APS/123-QED}
\title{Phase Reduction Method for Strongly Perturbed Limit-Cycle Oscillators}
\author{Wataru Kurebayashi}
\email{kurebayashi.w.aa@m.titech.ac.jp}
\author{Sho Shirasaka}
\author{Hiroya Nakao}
\affiliation{%
Graduate School of Information Science and Engineering,
Tokyo Institute of Technology, O-okayama 2-12, Meguro, Tokyo 152-8552, Japan
}%
\date{\today}


\begin{abstract}
The phase reduction method for limit-cycle oscillators subjected to weak perturbations has significantly contributed
to  theoretical investigations of rhythmic phenomena.
We here propose a generalized phase reduction method that is also applicable to strongly perturbed limit-cycle oscillators.
{The fundamental assumption of our method is} that the perturbations can be decomposed into a large-amplitude component varying slowly as compared to the amplitude relaxation time
and remaining weak fluctuations{.  Under this assumption,} we introduce a generalized phase parameterized by the slowly varying large-amplitude component
and derive a closed equation for the generalized phase describing the oscillator dynamics.
The proposed method enables us to explore a broader class of rhythmic phenomena,
in which the shape and frequency  of the oscillation may vary largely because of the perturbations.
We illustrate our method by analyzing the synchronization dynamics of limit-cycle oscillators driven by strong periodic signals.
It is shown that the proposed method accurately predicts the synchronization properties of the oscillators, 
while the conventional method does not.
\end{abstract}



\maketitle


Rhythmic phenomena are ubiquitous in nature and of great interest
in many fields of science and technology, including chemical reactions, neural networks, genetic circuits, lasers, and structural
vibrations~\cite{GBT,COWT,SUCNS,WCNN,LCOSync,Quorum,MFN,Brown}.
These rhythmic phenomena often result from complex interactions among individual rhythmic elements, typically modeled as limit-cycle oscillators.
In analyzing such systems, the phase reduction method~\cite{GBT,COWT,SUCNS,WCNN,MFN,Brown} has been widely used and considered an essential tool.
It systematically approximates the high-dimensional dynamical equation of a perturbed limit-cycle oscillator
by a one-dimensional reduced {\it phase equation}, with just a single {\it phase variable} $\theta$ representing the oscillator state.

A fundamental assumption of the conventional phase reduction method is that the applied perturbation is sufficiently weak;
hence, the shape and frequency of the limit-cycle orbit remain almost unchanged.
However, this assumption hinders the applications of the method to strongly perturbed oscillators because the shapes and
frequencies of their orbits can significantly differ from those in the unperturbed cases.
Indeed, strong coupling can destabilize synchronized states of
oscillators that are stable in the weak coupling limit~\cite{Aronson+Bressloff}.
The effect of strong coupling can further lead to nontrivial collective dynamics such as
quorum-sensing transition~\cite{Quorum}, amplitude death and bistability~\cite{Aronson+Bressloff}, and collective chaos~\cite{Hakim+Nakagawa}.
Although not all of these collective phenomena are the subject of discussion in this study, our formulation will give an insight into
a certain class of them, e.g., bistability between phase-locked and drifting states~\cite{Aronson+Bressloff}.
The assumption of weak perturbations can also be an obstacle to modeling real-world systems, which are often subjected to strong
perturbations.

Although the phase reduction method has recently been extended to stochastic~\cite{StochasticPRM}, delay-induced~\cite{DelayPRM},
and collective oscillations~\cite{CollectivePRM}, these extensions are still limited to the weakly perturbed regime.
To analyze a broader class of synchronization phenomena exhibited by strongly driven or interacting oscillators,
the conventional theory should be extended.
This Letter proposes an extension of the phase reduction method
to strongly perturbed limit-cycle oscillators,
which enables us to derive a simple generalized phase equation that quantitatively describes their dynamics.
We use it to analyze the synchronization dynamics
of limit-cycle oscillators subjected to strong periodic forcing,
which cannot be treated appropriately by the conventional method.


We consider a limit-cycle oscillator
whose dynamics depends on a time-varying parameter
$\bm{I}(t)$ $=$ $[I_1(t),$ $\ldots,$ $I_m(t)]^{\top}$ $\in\mathbb{R}^m$ representing general perturbations,
 described by
\begin{eqnarray}
 \dot{\bm{X}}(t)=\bm{F}(\bm{X}(t),\bm{I}(t)),
	\label{eq. model0}
\end{eqnarray}
where $\bm{X}(t)$ $=$ $[X_1(t),$ $\ldots,$ $X_n(t)]^{\top}$ $\in\mathbb{R}^n$ is the oscillator state
and $\bm{F}(\bm{X},\bm{I})$ $=$ $[F_1 (\bm{X},\bm{I}),$ $\ldots,$ $F_n (\bm{X},\bm{I}) ]^{\top}$ $\in\mathbb{R}^n$
is an $\bm{I}$-dependent vector field representing the oscillator dynamics.
For example, $\bm{X}$ and $\bm{I}$ can represent the state of a periodically firing neuron and the injected current, respectively \cite{WCNN,Brown}.
In this Letter, we introduce a generalized phase $\theta$, which depends on the parameter $\bm{I}(t)$, of the oscillator.
In defining the phase $\theta$, we require that 
the oscillator state $\bm{X}(t)$ can be accurately approximated by using $\theta(t)$ with sufficiently small error,
and that $\theta(t)$ increases at a constant frequency when the parameter $\bm{I}(t)$ remains constant.
The former requirement is a necessary condition for the phase reduction, i.e., for deriving a closed equation for the generalized phase,
and the latter enables us to derive an analytically tractable phase equation.

To define such $\theta$, we suppose that $\bm{I}$ is constant until further notice.
We assume that Eq.~(\ref{eq. model0}) possesses a family of stable limit-cycle solutions with period $T(\bm{I})$
and frequency  $\omega(\bm{I}) := 2\pi / T({\bm I})$ for ${\bm I} \in A$, where $A$ is an open subset
of $\mathbb{R}^{m}$ (e.g., an interval between two bifurcation points).
An oscillator state on the limit cycle with parameter ${\bm I}$ can be
parameterized by a phase $\theta\in[0,2\pi)$ as ${\bm X}_{0}(\theta, {\bm I})=[X_{0,1}(\theta,\bm{I}),\ldots,X_{0,n}(\theta,\bm{I})]^{\top}$.
Generalizing the conventional phase reduction method~\cite{GBT,COWT,SUCNS,WCNN,MFN}, we define the phase $\theta$ such that,
as the oscillator state ${\bm X}(t) = {\bm X}_{0}(\theta(t), {\bm I})$ evolves along the limit cycle, the corresponding phase $\theta(t)$
increases at a constant frequency $\omega({\bm I})$ as $\dot{\theta}(t) = \omega({\bm I})$ {\it for each} ${\bm I}\in A$.
We assume that ${\bm X}_{0}(\theta, {\bm I})$ is continuously differentiable with respect to $\theta\in[0,2\pi)$ and ${\bm I} \in A$.

\begin{figure*}[tbp]
 \begin{center}
	 \includegraphics[width=160mm]{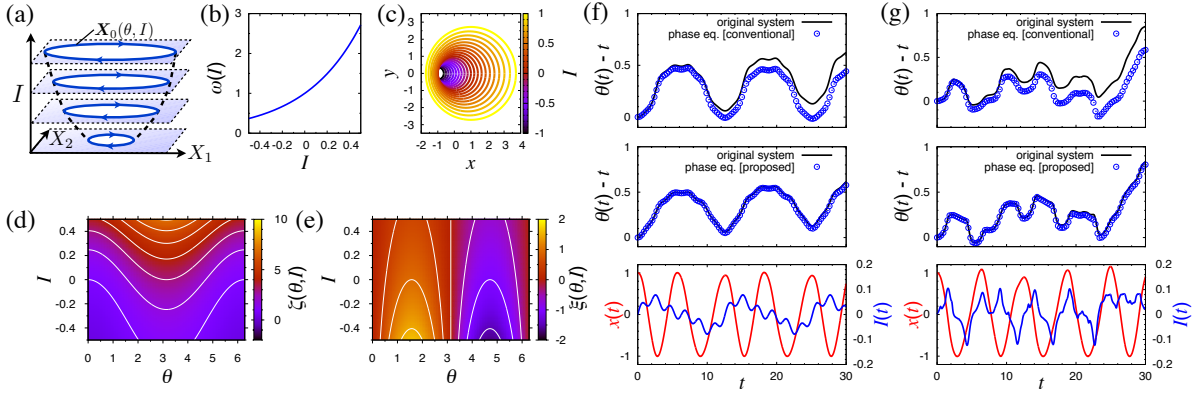}
	\end{center}
 \caption{(Color online) Phase dynamics of
 a modified Stuart-Landau oscillator.
 (a) A schematic diagram of
 the extended phase space $\mathbb{R}^n\times A$
 with $n=2$ and $m=1$.
 (b) Frequency $\omega(I)$.
 (c) $I$-dependent stable limit-cycle solutions $\bm{X}_0(\theta,I)$.
 (d), (e) Sensitivity functions $\zeta(\theta,I)$ and $\xi(\theta,I)$.
 (f), (g) Time series of the phase $\theta(t)$ of the oscillator driven by (f) a periodically varying parameter $I^{(1)}(t)$
 or (g) a chaotically varying parameter $I^{(2)}(t)$.
 For each of these cases, results of the conventional (top panel) and proposed (middle panel) methods are shown.
 Evolution of the conventional phase $\tilde{\theta}(t)=\Theta(\bm{X}(t),\bm{q}_c)$
 and the generalized phase $\theta(t)=\Theta(\bm{X}(t),\bm{q}(\epsilon t))$
 measured from the original system (lines) is compared with that of the conventional and generalized phase equations (circles).
 Time series of the state variable $x(t)$ (red) and time-varying parameter $I(t)$ (blue) are also depicted (bottom panel).
 {The periodically varying parameter is given by $I^{(1)}(t) = q^{(1)}(\epsilon t) + \sigma p^{(1)}(t)$
 with $q^{(1)}(\epsilon t)=0.05\sin(0.5t)+0.02\sin(t)$ and $\sigma p^{(1)}(t)=0.02\sin(3t)$,
 and the chaotically varying parameter is given by $I^{(2)}(t) = q^{(2)}(\epsilon t) + \sigma p^{(2)}(t)$
 with $q^{(2)}(\epsilon t)=0.007L_1(0.3t)$ and $\sigma p^{(2)}(t)=0.001L_2(t)$,
 where $L_1(t)$ and $L_2(t)$ are independently generated time series of the variable $x$ of the chaotic Lorenz equation~\cite{SUCNS},
 $\dot{x}=10(y-x)$, $\dot{y}=x(28-z)-y$, and $\dot{z}=xy-8z/3$.}}
 \label{fig. 0}
\end{figure*}

We consider an extended phase space $\mathbb{R}^n\times A$, as depicted schematically in Fig.~\ref{fig. 0} (a).
We define $C$ as a cylinder formed  by the family of limit cycles $(\bm{X}_0(\theta, \bm{I}), \bm{I})$ for $\theta\in[0,2\pi)$ and $\bm{I}\in A$,
and define $U\subset\mathbb{R}^n\times A$ as a neighborhood of $C$.
For each ${\bm I}$, we assume that any orbit starting from an arbitrary point $({\bm X}, {\bm I})$ in $U$ asymptotically converges to
the limit cycle ${\bm X}_0(\theta, {\bm I})$ on $C$.
We can then extend the definition of the phase into $U$, as in the conventional method~\cite{GBT,COWT,SUCNS,WCNN,MFN},
by introducing the {\it asymptotic phase} and {\it isochrons} around the limit cycle for each ${\bm I}$.
Namely, we can define a generalized {\it phase function} $\Theta(\bm{X},\bm{I})\in[0,2\pi)$ of $({\bm X}, {\bm I}) \in U$
such that $\Theta(\bm{X},\bm{I})$ is continuously differentiable with respect to ${\bm X}$ and ${\bm I}$, and
$\frac{\partial\Theta(\bm{X},\bm{I})}{\partial\bm{X}} \cdot \bm{F}(\bm{X},\bm{I}) =\omega(\bm{I})$
holds everywhere in $U$,
where $\frac{\partial\Theta}{\partial\bm{X}} = [\frac{\partial\Theta}{\partial X_1},$ $\ldots,$ $\frac{\partial\Theta}{\partial X_n}]^{\top}$ $\in\mathbb{R}^n$
is the gradient of $\Theta(\bm{X},\bm{I})$ with respect to $\bm{X}$ and the dot $(\cdot)$ denotes an inner product.
This $\Theta(\bm{X},\bm{I})$ is a straightforward generalization of the conventional asymptotic phase~\cite{GBT,COWT,SUCNS,WCNN,MFN}
and guarantees that the phase of any orbit ${\bm X}(t)$ in $U$ always increases with a constant frequency as
$\dot{\Theta}(\bm{X}(t),\bm{I}) = \omega({\bm I})$ at each ${\bm I}$.
For any oscillator state on $C$, $\Theta({\bm X}_{0}(\theta, {\bm I}), {\bm I}) = \theta$ holds.
In general, the origin of the phase can be arbitrarily defined for each ${\bm I}$
as long as it is continuously differentiable with respect to ${\bm I}$.
The assumptions that $\bm{X}_0(\theta,\bm{I})$ and $\Theta(\bm{X},\bm{I})$
are continuously differentiable can be further relaxed for a certain class of oscillators, such as those considered in~\cite{Izhikevich_Relax}.

Now suppose that the parameter ${\bm I}(t)$ varies with time.
To define $\theta$ that approximates the oscillator state with sufficiently small error,
we assume that $\bm{I}(t)$ can be decomposed into a slowly varying component $\bm{q}(\epsilon t)\in A$
and remaining weak fluctuations $\sigma\bm{p}(t)\in\mathbb{R}^m$ as ${\bm I}(t) = \bm{q}(\epsilon t) + \sigma\bm{p}(t)$.
Here, the parameters $\epsilon$ and $\sigma$ are assumed to be sufficiently small
so that $\bm{q}(\epsilon t)$ varies  slowly as compared to the relaxation time of a perturbed orbit to the cylinder $C$ of the limit cycles,
which we assume to be $O(1)$ without loss of generality,
and the oscillator state ${\bm X}(t)$ always remains in a close neighborhood of ${\bm X}_0(\theta, {\bm q}(\epsilon t))$ on $C$,
i.e., $\bm{X}(t) = \bm{X}_0(\theta(t),\bm{q}(\epsilon t)) + O(\epsilon,\sigma)$ holds (see Supplementary Information).
We also assume that $\bm{q}(\epsilon t)$ is continuously differentiable with respect to $t\in\mathbb{R}$.
Note that the slow component ${\bm q}(\epsilon t)$ itself does not need to be small.



Using the phase function $\Theta(\bm{X},\bm{I})$, we introduce a generalized phase $\theta(t)$ of the limit-cycle oscillator~(\ref{eq. model0})
as $\theta(t)=\Theta(\bm{X}(t),\bm{q}(\epsilon t))$.
This definition guarantees that $\theta(t)$ increases at a constant frequency when $\bm{I}(t)$ remains constant,
and leads to a closed equation for $\theta(t)$.
Expanding Eq.~(\ref{eq. model0}) in $\sigma$ as
$\dot{\bm{X}}(t) =\bm{F}(\bm{X},\bm{q}(\epsilon t))
+\sigma \bm{G}(\bm{X},\bm{q}(\epsilon t))\bm{p}(t)+O(\sigma^2)$
and using the chain rule, we can derive
$\dot{\theta}(t) = \omega(\bm{q}(\epsilon t)) + \sigma\frac{\partial\Theta(\bm{X},\bm{I})}{\partial\bm{X}}
 |_{(\bm{X}(t),\bm{q}(\epsilon t))} \cdot \bm{G}(\bm{X},\bm{q}(\epsilon t))\bm{p}(t)
+\epsilon\frac{\partial\Theta(\bm{X},\bm{I})}{\partial\bm{I}}
 |_{(\bm{X}(t),\bm{q}(\epsilon t))} \cdot\dot{\bm{q}}(\epsilon t) + O(\sigma^2),$
where $\bm{G}(\bm{X},\bm{I})\in\mathbb{R}^{n\times m}$ is a matrix whose ($i,j$)-th element is
given by $\frac{\partial F_i(\bm{X},\bm{I})}{\partial I_j}$,
$\frac{\partial\Theta}{\partial\bm{I}} = [\frac{\partial\Theta}{\partial I_1},$ $\ldots,$ $\frac{\partial\Theta}{\partial I_{m}}]^{\top}$ $\in\mathbb{R}^{m}$
is the gradient of $\Theta(\bm{X},\bm{I})$ with respect to ${\bm I}$, and $\dot{\bm{q}}(\epsilon t)$ denotes $\frac{d\bm{q}(\epsilon t)}{d(\epsilon t)}$.

To obtain a closed equation for $\theta$, we use the lowest-order approximation in $\sigma$ and $\epsilon$,
i.e., ${\bm X}(t) = {\bm X}_0(\theta(t),{\bm q}(\epsilon t)) + O(\epsilon,\sigma)$.
Then, by defining a {\it phase sensitivity function}
$\bm{Z}(\theta,\bm{I})=\frac{\partial\Theta(\bm{X},\bm{I})}{\partial\bm{X}}|_{(\bm{X}_0(\theta,\bm{I}), {\bm I})}\in\mathbb{R}^{n}$
and two other sensitivity functions $\bm{\zeta}(\theta,\bm{I})=\bm{G}^{\top}(\bm{X}_0(\theta,\bm{I}),\bm{I})\bm{Z}(\theta,\bm{I}) \in\mathbb{R}^{m}$ and
$\bm{\xi}(\theta,\bm{I})= \frac{\partial\Theta(\bm{X},\bm{I})}{\partial\bm{I}}|_{(\bm{X}_0(\theta, \bm{I}), {\bm I})} \in\mathbb{R}^{m}$,
we can obtain a closed equation for the oscillator phase $\theta(t)$ as
\begin{eqnarray}
 \dot{\theta}(t) &=& \omega(\bm{q}(\epsilon t)) + \sigma \bm{\zeta}(\theta,\bm{q}(\epsilon t))\cdot\bm{p}(t) \cr
	&& + \epsilon \bm{\xi}(\theta,\bm{q}(\epsilon t)) \cdot \dot{\bm{q}}(\epsilon t) + O(\sigma^2,\epsilon^2,\sigma\epsilon),
	\label{eq. reduce2}
\end{eqnarray}
which is a generalized phase equation that we propose in this study.
The first three terms in the right-hand side of Eq.~(\ref{eq. reduce2}) represent the instantaneous frequency of the oscillator,
the phase response to the weak fluctuations $\sigma {\bm p}(t)$, and the phase response to deformation of the limit-cycle orbit
caused by the slow variation in ${\bm q}(\epsilon t)$, respectively, all of which depend on the slowly varying component $\bm{q}(\epsilon t)$.

To address the validity of Eq.~(\ref{eq. reduce2}) more precisely, let $\lambda(\bm{I}) \; (>0) $ denote the absolute value of the second largest Floquet exponent of the
oscillator for a fixed $\bm{I}$, which characterizes the amplitude relaxation timescale of the oscillator ($\approx 1/\lambda(\bm{I})$).
As argued in Supplementary Information, we can show that the error terms in Eq.~(\ref{eq. reduce2})
remain sufficiently small when $\sigma/\lambda(\bm{q}(\epsilon t)) \ll 1$ and $\epsilon/\lambda(\bm{q}(\epsilon t))^2 \ll 1${,
namely, when the orbit of the oscillator relaxes to the cylinder $C$ sufficiently faster than the variations in $\bm{q}(\epsilon t)$.}

Note that if we define the phase variable as $\tilde{\theta}(t)=\Theta(\bm{X}(t),\bm{q}_c)$ with some constant $\bm{q}_c$
instead of $\theta(t)=\Theta(\bm{X}(t),\bm{q}(\epsilon t))$,
$\tilde{\theta}(t)$ gives the conventional phase.
Then, we obtain the conventional phase equation
$\dot{\tilde{\theta}}(t)=\omega_c +\sigma\bm{\zeta}_c(\tilde{\theta})\cdot\bm{p}(t) +O(\sigma^2)$
with $\bm{q}(\epsilon t)=\bm{q}_c$ and $\sigma\bm{p}(t)=\bm{I}(t)-\bm{q}_c$.
Here, $\omega_c:=\omega(\bm{q}_c)$ is a natural frequency,
$\bm{\zeta}_c(\tilde{\theta}) =\bm{\zeta}(\tilde{\theta}, \bm{q}_c) =\bm{G}(\bm{X}_0(\tilde{\theta}, \bm{q}_c),\bm{q}_c)^{\top} \bm{Z}(\tilde{\theta}, \bm{q}_c)$,
and $\bm{Z}(\tilde{\theta}, \bm{q}_c)$ is the conventional phase sensitivity function at $\bm{I}=$ $\bm{q}_c$~\cite{COWT}.
This equation is valid only when $\sigma/\lambda(\bm{q}_c)\ll1$ (i.e., $||\bm{I}(t)-\bm{q}_c||/\lambda(\bm{q}_c)\ll1$).
{By using the near-identity transformation~\cite{Keener},
we can show that the conventional equation is actually a low-order approximation of the generalized equation~(\ref{eq. reduce2})
(see Sec. III of Supplementary Information).}


In practice, we need to calculate $\bm{\zeta}(\theta,\bm{I})$ and $\bm{\xi}(\theta,\bm{I})$ numerically
from mathematical models or estimate them through experiments.
We can show that the following relations hold
(See Supplementary Information for the derivation):
\begin{eqnarray}
 \bm{\xi}(\theta,\bm{I}) &=&
	-\frac{\partial\bm{X}_0(\theta, \bm{I})}{\partial \bm{I}}^{\top} \bm{Z}(\theta,\bm{I}),
	\label{eq. def Y 2} \\
 \bm{\xi}(\theta,\bm{I})&=&\bm{\xi}(\theta_0,\bm{I})
	-\frac{1}{\omega(\bm{I})} \int_{\theta_0}^{\theta} [\bm{\zeta}(\theta',\bm{I})-\bar{\bm{\zeta}}(\bm{I})]d\theta',
	\label{eq. def Y 3}
\end{eqnarray}
\vspace{-0.8cm}
\begin{eqnarray}
 \bar{\bm{\zeta}}(\bm{I})&:=&\frac{1}{2\pi}\int_0^{2\pi}\bm{\zeta}(\theta,\bm{I})d\theta
	=\frac{d\omega(\bm{I})}{d\bm{I}},
	\label{eq. def Y 4}
\end{eqnarray}
where $\frac{\partial\bm{X}_0(\theta,\bm{I})}{\partial\bm{I}} \in\mathbb{R}^{n\times m}$ is a matrix whose ($i,j$)-th element is given by
$\frac{\partial X_{0,i}(\theta,\bm{I})}{\partial I_j}$, $\theta_0\in[0,2\pi)$ is a constant, and
$\bar{\bm{\zeta}}(\bm{I})$ is the average of $\bm{\xi}(\theta,\bm{I})$ with respect to $\theta$ over one period of oscillation.
From mathematical models of limit-cycle oscillators, $\bm{Z}(\theta,\bm{I})$ can be obtained numerically by
the adjoint method for each ${\bm I}$~\cite{MFN,Brown}, and then $\bm{\zeta}(\theta,\bm{I})$ and $\bm{\xi}(\theta,\bm{I})$
can be computed from $\bm{\zeta}(\theta,\bm{I})=\bm{G}^{\top}(\bm{X}_0(\theta,\bm{I}),\bm{I})\bm{Z}(\theta,\bm{I})$
and Eqs.~(\ref{eq. def Y 2}) and (\ref{eq. def Y 3}).
Experimentally, $\bm{Z}(\theta,\bm{I})$ and $\bm{\zeta}(\theta,\bm{I})$ can be measured
by applying small impulsive perturbations to ${\bm I}$,
while $\bm{\xi}(\theta,\bm{I})$ can be obtained by applying small stepwise perturbations to ${\bm I}$.


To test the validity of the generalized phase equation~(\ref{eq. reduce2}),
we introduce an analytically tractable model, a modified Stuart-Landau (MSL) oscillator (see \cite{MSL} and Fig.~\ref{fig. 0} for the definition and details).
We numerically predict the phase $\theta(t)$ of a strongly perturbed MSL oscillator by both conventional and generalized phase equations,
and compare them with direct numerical simulations of the original system.
In applying the conventional phase reduction, we set $q_c=\langle I(t)\rangle_t$, where $\langle\cdot\rangle_t$ denotes the time average.
In Fig.~\ref{fig. 0}, we can confirm that the generalized phase equation~(\ref{eq. reduce2}) accurately predicts
the generalized phase $\Theta(\bm{X}(t),q(\epsilon t))$ of the original system,
while the conventional phase equation does not well predict the conventional phase $\Theta(\bm{X}(t),q_c)$ because of large variations in $I(t)$.


As an application of the generalized phase equation (\ref{eq. reduce2}), we analyze {\it $k:l$ phase locking}~\cite{HighOrderLocking}
of the system~(\ref{eq. model0}) to a periodically varying parameter $\bm{I}(t)$ with period $T_I$ and frequency $\omega_I$,
in which the frequency tuning ($l\langle\dot{\theta}\rangle_t=k\omega_I$) occurs.
Although the averaging approximation~\cite{averaging}
for the phase difference $\tilde{\psi}(t) = l\theta(t)-k\omega_I$ is generally used to analyze the phase locking~\cite{COWT,HighOrderLocking},
we cannot directly apply it in the present case because the frequency $\omega(\bm{q}(\epsilon t))$ can vary largely with time.
Thus, generalizing the conventional definition, we introduce the phase difference as
$\psi(t)=l\theta(t)-k\omega_I t-lh(t)$ with an additional term $-lh(t)$
to remove the large periodic variations in $\psi(t)$ due to $\omega(\bm{q}(\epsilon t))$, where
$h(t)$ is a $T_I$-periodic function defined as $h(t)=\int_0^t [ \omega(\bm{q}(\epsilon t'))-T_I^{-1}\int_0^{T_I}\omega(\bm{q}(\epsilon t))dt ] dt'$.
By virtue of this term, temporal variations in $\dot{\psi}$ remain of the order $O(\epsilon,\sigma)$,
i.e., $|\dot{\psi}| \ll 1$, which enables us to apply the averaging approximation to $\psi$.

Introducing a small parameter $\nu$ representing the magnitude of variations in $\psi$, one can derive a dynamical equation for $\psi$ as
$\dot{\psi}(t) = \nu f(\psi,t),$
where $\nu f(\psi,t)=lg(\psi/l+k\omega_I t/l+h(t),t)
-k\omega_I-l\dot{h}(t)$
and $g(\theta,t)$ denotes
the right-hand side of Eq.~(\ref{eq. reduce2}).
Using  first- and second-order averaging~\cite{averaging}, we can introduce slightly deformed phase differences $\psi_{1,2}$
satisfying $\psi_{1,2}(t)=\psi(t)+O(\nu)$ and obtain the first- and second-order averaged equations,
$\dot{\psi}_1(t)=\nu \bar{f}_1(\psi_1) +O(\nu^2)$
and
$\dot{\psi}_2(t)=\nu \bar{f}_1(\psi_2) + \nu^2\bar{f}_2(\psi_2) +
O(\nu^3),$
where $\bar{f}_1(\psi)$ and $\bar{f}_2(\psi)$ are given by
$\bar{f}_1(\psi)=(lT_I)^{-1}\int_0^{lT_I}f(\psi,t)dt,$
$\bar{f}_2(\psi)=(lT_I)^{-1}\int_0^{lT_I}
	[u(\psi,t) \frac{\partial f(\psi,t)}{\partial\psi}
	-\bar{f}_1(\psi) \frac{\partial u(\psi,t)}{\partial\psi}]dt,$
and $u(\psi,t)=\int_0^t[f(\psi,t')-\bar{f}_1(\psi)]dt'$.
These averaged equations can be considered autonomous by neglecting the $O(\nu^2)$ and $O(\nu^3)$ terms, respectively.
Averaged equations for the conventional phase equation can be derived similarly.
Thus, if the averaged equation has a stable fixed point, $k:l$ phase locking is expected to occur.
As demonstrated below, the first-order averaging of the generalized phase equation already predicts qualitative features of the phase-locking dynamics,
while the second-order averaging gives more precise results when the parameter $\bm{I}(t)$ varies significantly.



\begin{figure*}[tbp]
 \begin{center}
	\includegraphics[width=160mm]{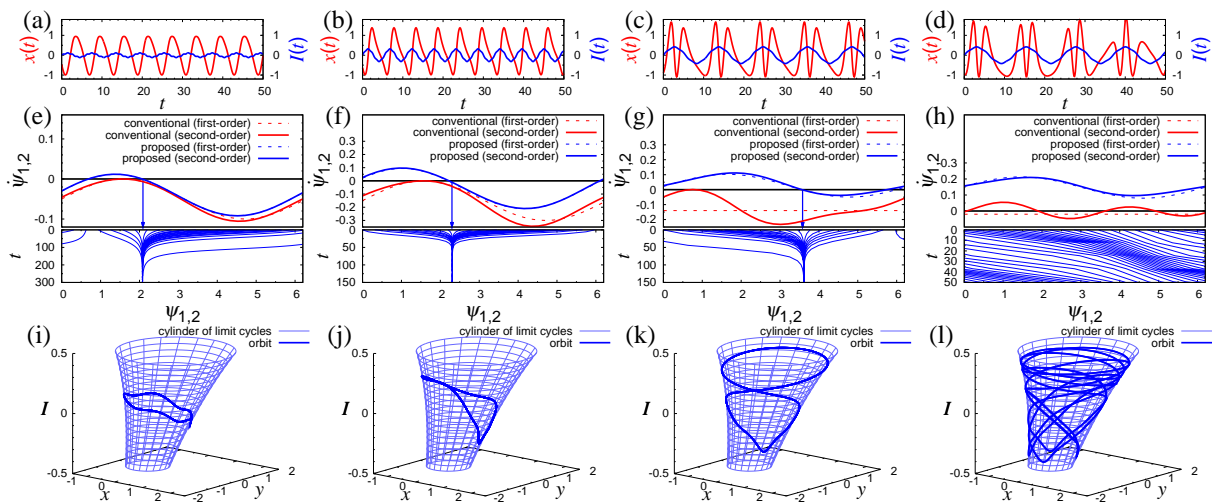}
 \end{center}
 \caption{(Color online) Phase locking of
 the modified Stuart-Landau oscillator.
 Four types of periodically varying parameters
 $I^{(j)}$ ($j=3,4,5,6$) are applied, which lead to
 $1:1$ phase locking to
 $I^{(3)}(t)$ [(a), (e), and (i)],
 $1:1$ phase locking to
 $I^{(4)}(t)$ [(b), (f), and (j)],
 $1:2$ phase locking to
 $I^{(5)}(t)$ [(c), (g), and (k)],
 and
 failure of phase locking to
 $I^{(6)}(t)$ [(d), (h), and (l)].
 (a)--(d) Time series of the state variable $x(t)$
 of a periodically driven oscillator (red)
 and periodic external forcing (blue).
 (e)--(h) Dynamics of the phase difference $\psi_{1,2}$
 with an arrow representing a stable fixed point
 (top panel)
 and time series of $\psi_{1,2}$ with 20 different initial states
 (bottom panel).
 (i)--(l) Orbits of a periodically driven oscillator (blue)
 on the cylinder of the limit cycles (light blue)
 plotted in the extended phase space.
 {The parameter $I^{(j)}(t)$ is given by
 $I^{(j)}(t) = q^{(j)}(\epsilon t) + \sigma p^{(j)} (t)$,
 $q^{(j)}(\epsilon t)=\alpha^{(j)}\sin(\omega_I^{(j)} t)$
 and $\sigma p^{(j)} (t)=0.02\sin(5\omega_I^{(j)} t)$
 with $\alpha^{(3,4,5,6)}$ = 0.1, 0.3, 0.4, 0.4,
 and $\omega_I^{(3,4,5,6)}$ = 1.05, 1.10, 0.57, 0.51.}}
 \label{fig. 1}
\end{figure*}

As an example, we use the MSL oscillator
and investigate their phase locking to periodic forcing.
Figure~\ref{fig. 1} shows the results of the numerical simulations.
We apply four types of periodically varying parameters and predict if the oscillator exhibits either $1:1$ or $1:2$ phase locking
to the periodically varying parameter $q(\epsilon t)$ (small fluctuation $\sigma p(t)$ is also added for completeness).
We derive averaged equations for the phase differences $\psi_{1,2}$ using the proposed and conventional methods,
and compare the results with direct numerical simulations of the MSL oscillator.
We find that our new method correctly predicts the stable phase-locking point already at first-order averaging,
while the conventional method does not.
In particular, the conventional method can fail to predict whether phase locking takes place or not, as shown in
Figs.~\ref{fig. 1}~(g) and (h), even after the second-order averaging.
{In this case, the exponential dependence of the frequency $\omega(I)$ on the parameter $I$ is the main cause of the breakdown of the conventional method (see Sec. III of Supplementary Information for a discussion).}
Typical trajectories of $[x(t),$ $y(t),$ $q(\epsilon t)]^{\top}$ are plotted
on the cylinder $C$ of limit cycles in the extended phase space $[x,y,I]^{\top}$,
which shows that the oscillator state migrates over $C$ synchronously with the periodic forcing.
The trajectories are closed when phase locking occurs.


In summary, we proposed a generalized phase reduction method that enables us to theoretically explore a broader class of
strongly perturbed limit-cycle oscillators.
Although still limited to slowly varying perturbations with weak fluctuations, our method avoids the assumption of weak perturbations,
which has been a major obstacle in applying the conventional phase reduction method to real-world phenomena.
It will therefore facilitate further theoretical investigations of nontrivial synchronization phenomena of strongly
perturbed limit-cycle oscillators~\cite{Aronson+Bressloff,Hakim+Nakagawa}.
As a final remark, we point out that a phase equation similar to Eq.~(\ref{eq. reduce2}) has been postulated
in a completely different context, to analyze the {\it geometric phase} in dissipative dynamical systems~\cite{Kepler}.
This formal similarity may provide an interesting possibility of understanding
synchronization dynamics of strongly perturbed oscillators from a geometrical viewpoint.


Financial support by JSPS KAKENHI (25540108, 22684020),
CREST Kokubu project of JST, and FIRST Aihara project of JSPS are gratefully acknowledged.

\newpage

\def\EQPhaseEq{2}
\def\EQPropZZetaXiA{3} 
\def\EQPropZZetaXiB{4} 
\def\EQPropOmegZeta{5} 

\section*{Supplemental Material}

\section{Derivation of the generalized phase equation}

In this section, we give a  detailed derivation of
the generalized phase equation (\EQPhaseEq) in the main article,
which takes into account the effect of amplitude relaxation
of the oscillator state to the cylinder of limit cycles $C$.
Our aim is to evaluate the order
of error terms in the generalized phase equation (\EQPhaseEq).
Our argument here is based on a formulation similar to Ref.~\cite{Goldobin}
by Goldobin {\it et al.}, in which the effect of colored noise on
limit-cycle oscillators is analyzed and an effective phase equation
that accurately describes the oscillator state is derived
by incorporating the effect of amplitude relaxation of
the oscillator state to the unperturbed limit-cycle orbit.

As in the main article,
we consider a limit-cycle oscillator
whose dynamics depends on a time-varying parameter $\bm{I}(t)$
representing general perturbations, described by
\begin{eqnarray}
 \dot{\bm{X}}(t)=\bm{F}(\bm{X}(t),\bm{I}(t)).
	\label{eq. model0}
\end{eqnarray}
For simplicity, we assume that the state variable $\bm{X}(t)$ is two-dimensional ($n=2$),
but the formulation can be straightforwardly extended to  higher-dimensional cases.

Suppose that the parameter $\bm{I}$ is constant for the moment.
As explained in the main article, we introduce
an extended phase space $\mathbb{R}^n\times A$ and define a generalized phase $\theta$
and amplitude $r$ as functions of $(\bm{X},\bm{I})$ in $U$.
Here, $r$ gives the distance of the oscillator state ${\bm X}$
from the unperturbed stable limit cycle ${\bm X}_{0}(\theta, {\bm I})$.
For each constant value of $\bm{I} \in A$,
as argued in the Supplementary Information of Ref.~\cite{Goldobin},
we can define a phase $\theta=\Theta(\bm{X},\bm{I})$
and an amplitude $r=R(\bm{X},\bm{I})$ such that
\begin{eqnarray}
 \frac{\partial\Theta(\bm{X},\bm{I})}{\partial\bm{X}}\cdot
\bm{F}(\bm{X},\bm{I})&=&\omega(\bm{I}), \label{eq. theta def}\\
 \frac{\partial R(\bm{X},\bm{I})}{\partial\bm{X}}\cdot
\bm{F}(\bm{X},\bm{I})&=&-\lambda(\bm{I})R(\bm{X},\bm{I}), \label{eq. r def}
\end{eqnarray}
where $\lambda(\bm{I})$ is the absolute value of the second Floquet exponent
of Eq.~(\ref{eq. model0}) for each $\bm{I}$.
We further assume that
$\Theta(\bm{X},\bm{I})$ and $R(\bm{X},\bm{I})$
are continuously differentiable
with respect to $\bm{X}$ and $\bm{I}$.
Equations~(\ref{eq. theta def}) and (\ref{eq. r def}) guarantee
that
\begin{align}
\dot{\theta}=\omega(\bm{I}),
\quad
\dot{r}=-\lambda(\bm{I})r
\end{align}
always hold for each $\bm{I}$.
In the absence of perturbations, the amplitude $r = R({\bm X}, {\bm I})$ decays to $0$ exponentially, and the phase $\theta = \Theta({\bm X}, {\bm I})$ increases constantly.

Now we suppose that the parameter $\bm{I}(t)$ can vary with time.
As explained in the main article, we decompose the parameter $\bm{I}(t)$
into a slowly varying component $\bm{q}(\epsilon t)$
and remaining weak fluctuations $\sigma\bm{p}(t)$ as $\bm{I}(t) = {\bm q}(\epsilon t) + \sigma \bm{p}(t)$.
We define a phase $\theta(t)$
and an amplitude $r(t)$ of the oscillator as follows:
\begin{eqnarray}
 \theta(t)&=&\Theta(\bm{X}(t),\bm{q}(\epsilon t)),\\
 r(t)&=&R(\bm{X}(t),\bm{q}(\epsilon t)).
\end{eqnarray}
Since $\Theta(\bm{X},\bm{I})$
and $R(\bm{X},\bm{I})$ are continuously differentiable
with respect to $\bm{X}$ and $\bm{I}$,
we can derive the dynamical equations
for $\theta(t)$ and $r(t)$ as
\begin{eqnarray}
 \dot{\theta}&=& 
	\left.\frac{\partial\Theta(\bm{X},\bm{I})}{\partial\bm{X}}
	\right|_{(\bm{X},\bm{q}(\epsilon t))}
	\cdot\frac{d\bm{X}(t)}{dt}
	+\left.\frac{\partial\Theta(\bm{X},\bm{I})}{\partial \bm{I}}
	\right|_{(\bm{X},\bm{q}(\epsilon t))}
	\cdot\frac{d\bm{q}(\epsilon t)}{dt}, \label{eq. theta2}\\
 \dot{r}&=&
	\left.\frac{\partial R(\bm{X},\bm{I})}{\partial\bm{X}}
	\right|_{(\bm{X},\bm{q}(\epsilon t))}
  \cdot\frac{d\bm{X}(t)}{dt}
	+\left.\frac{\partial R(\bm{X},\bm{I})}{\partial \bm{I}}
	\right|_{(\bm{X},\bm{q}(\epsilon t))}
	\cdot\frac{d\bm{q}(\epsilon t)}{dt}. \label{eq. r2}
	\label{eq. r exact2}
\end{eqnarray}
Plugging $\bm{I}(t)=\bm{q}(\epsilon t)+\sigma\bm{p}(t)$
into Eq.~(\ref{eq. model0}) and expanding it to the first order in
$\sigma$, we can derive
\begin{eqnarray}
 \dot{\bm{X}}=\bm{F}(\bm{X},\bm{q}(\epsilon t))
	+\sigma\bm{G}(\bm{X},\bm{q}(\epsilon t))\bm{p}(t)+O(\sigma^2),
	\label{eq. expand}
\end{eqnarray}
where the matrix ${\bm G}$ is defined in the main article.
Substituting Eqs.~(\ref{eq. theta def}), (\ref{eq. r def}),
and (\ref{eq. expand}) into Eqs.~(\ref{eq. theta2}) and (\ref{eq. r2}), we can obtain
\begin{align}
 \dot{\theta}&=
	\omega(\bm{q}(\epsilon t))
	+\sigma\left.\frac{\partial\Theta(\bm{X},\bm{I})}{\partial\bm{X}}
	\right|_{(\bm{X},\bm{q}(\epsilon t))}
	\cdot\bm{G}(\bm{X},\bm{q}(\epsilon t))\bm{p}(t)
	+\epsilon\left.\frac{\partial\Theta(\bm{X},\bm{I})}{\partial \bm{I}}
	\right|_{(\bm{X},\bm{q}(\epsilon t))}
	\cdot\dot{\bm{q}}(\epsilon t)+O(\sigma^2), \label{eq. theta3} \\
 \dot{r}&=
	-\lambda(\bm{q}(\epsilon t))r
	+\sigma\left.\frac{\partial R(\bm{X},\bm{I})}{\partial\bm{X}}
	\right|_{(\bm{X},\bm{q}(\epsilon t))}
  \cdot\bm{G}(\bm{X},\bm{q}(\epsilon t))\bm{p}(t)
 +\epsilon\left.\frac{\partial R(\bm{X},\bm{I})}{\partial \bm{I}}
	\right|_{(\bm{X},\bm{q}(\epsilon t))}
	\cdot\dot{\bm{q}}(\epsilon t)+O(\sigma^2), \label{eq. r3}
\end{align}
where $\dot{\bm{q}}(\epsilon t)$ denotes
$d\bm{q}(\epsilon t)/d(\epsilon t)$.
For simplicity of notation, we define
$\bm{\zeta}_{\theta}(\theta,r,\bm{I})\in\mathbb{R}^m$,
$\bm{\zeta}_r(\theta,r,\bm{I})\in\mathbb{R}^m$,
$\bm{\xi}_{\theta}(\theta,r,\bm{I})\in\mathbb{R}^m$ and
$\bm{\xi}_r(\theta,r,\bm{I})\in\mathbb{R}^m$, respectively, as
\begin{eqnarray}
\bm{\zeta}_{\theta}(\theta,r,\bm{I})&=&
 \left.\bm{G}(\bm{X},\bm{I})^{\top}
	\frac{\partial\Theta(\bm{X},\bm{I})}{\partial\bm{X}}
	\right|_{\bm{X}=\bm{X}(\theta,r,\bm{I})}, \label{eq. zeta1}\\
 \bm{\zeta}_r(\theta,r,\bm{I})&=&
	\left.\bm{G}(\bm{X},\bm{I})^{\top}
 \frac{\partial R(\bm{X},\bm{I})}{\partial\bm{X}}
 \right|_{\bm{X}=\bm{X}(\theta,r,\bm{I})}, \label{eq. zeta2} \\
 \bm{\xi}_{\theta}(\theta,r,\bm{I})&=&
	\left.\frac{\partial\Theta(\bm{X},\bm{I})}{\partial\bm{I}}
	 \right|_{\bm{X}=\bm{X}(\theta,r,\bm{I})}, \label{eq. xi1} \\
 \bm{\xi}_r(\theta,r,\bm{I})&=&
 \left.\frac{\partial R(\bm{X},\bm{I})}{\partial\bm{I}}
				\right|_{\bm{X}=\bm{X}(\theta,r,\bm{I})}, \label{eq. xi2}
\end{eqnarray}
where $\bm{X}(\theta,r,\bm{I})$
$\in \mathbb{R}^2$ represents an oscillator state with 
$\theta=\Theta(\bm{X},\bm{I})$, $r=R(\bm{X},\bm{I})$,
and parameter ${\bm I}$.
Using Eqs.~(\ref{eq. zeta1}), (\ref{eq. zeta2}),
(\ref{eq. xi1}), and (\ref{eq. xi2}),
we can rewrite Eqs.~(\ref{eq. theta3})
and (\ref{eq. r3}) as
\begin{eqnarray}
  \dot{\theta}&=&
	\omega(\bm{q}(\epsilon t))
	+\sigma\bm{\zeta}_{\theta}(\theta,r,\bm{I})
	\cdot\bm{p}(t)
	+\epsilon\bm{\xi}_{\theta}(\theta,r,\bm{I})
	\cdot\dot{\bm{q}}(\epsilon t)+O(\sigma^2), \label{eq. theta4} \\
 \dot{r}&=&
-\lambda(\bm{q}(\epsilon t))r
+\sigma\bm{\zeta}_r(\theta,r,\bm{I})
  \cdot\bm{p}(t)
	+\epsilon\bm{\xi}_r(\theta,r,\bm{I})
	\cdot\dot{\bm{q}}(\epsilon t)+O(\sigma^2). \label{eq. r4}
\end{eqnarray}
Note that $\bm{\zeta}_{\theta}(\theta,0,\bm{I})$ and $\bm{\xi}_{\theta}(\theta,0,\bm{I})$ are equivalent to the sensitivity functions
$\bm{\zeta}(\theta,\bm{I})$ and $\bm{\xi}(\theta,\bm{I})$ defined in the main article.
The functions $\bm{\zeta}_r(\theta,r,\bm{I})$ and $\bm{\xi}_r(\theta,r,\bm{I})$ represent sensitivities of the amplitude to the small fluctuations and to the slowly varying component of the applied perturbations, respectively.
In the main article, we also assumed that
 $\bm{q}(\epsilon t)$
varies sufficiently slowly as compared to
the relaxation time of perturbed orbits to $C$.
By using the absolute value of the Floquet exponent $\lambda(\bm{I})$
and the slowly varying component $\bm{q}(\epsilon t)$,
this assumption can be written as
\begin{eqnarray}
 \epsilon \ll \lambda(\bm{q}(\epsilon t)),
	\,\,\,{\rm or}\,\,\,
	\frac{\epsilon}{\lambda(\bm{q}(\epsilon t))}\ll1.
\end{eqnarray}

Now, we show that the following relation between the sensitivity functions for the amplitude holds:
\begin{align}
 \bm{\xi}_r(\theta,0,\bm{I})
 &= - \frac{1}{\omega(\bm{I})} \int_{0}^{\infty} e^{-\lambda(\bm{I})\phi/\omega(\bm{I})} \bm{\zeta}_r(\theta-\phi,0,\bm{I})d\phi
 \label{eq. xi r - xeta eta} \\
 &= - \frac{1}{\lambda(\bm{I})} \int_{0}^{\infty} e^{-s} \bm{\zeta}_r(\theta-\omega(\bm{I})s/\lambda(\bm{I}),0,\bm{I})ds,
 \label{eq. xi r - xeta eta2}
\end{align}
where we defined $s = \lambda({\bm I}) \phi / \omega({\bm I})$ in the second line.
{From Eq.~(\ref{eq. r def}),
\begin{align}
 \frac{\partial R(\bm{X},\bm{I})}{\partial\bm{X}} \cdot \bm{F}(\bm{X},\bm{I}) &= - \lambda(\bm{I}) R(\bm{X},\bm{I})
 \label{eq. def R func}
\end{align}
holds. We differentiate Eq.~(\ref{eq. def R func}) with respect to $\bm{I}$
and plug in $\bm{X}=\bm{X}_0(\theta,\bm{I})$. 
Then, from the left-hand side of Eq.~(\ref{eq. def R func}), we obtain
\begin{align}
 \left. \frac{\partial}{\partial\bm{I}} \left[ \frac{\partial R(\bm{X},\bm{I})}{\partial\bm{X}} \cdot \bm{F}(\bm{X},\bm{I}) \right] \right|_{\bm{X} = \bm{X}_0(\theta,\bm{I})}
 &= \left.\left[\frac{\partial}{\partial\bm{I}}\left(\frac{\partial R(\bm{X},\bm{I})}{\partial\bm{X}}\right)\right]^{\top}
 \bm{F}(\bm{X},\bm{I}) \right|_{\bm{X}=\bm{X}_0(\theta,\bm{I})}
 \cr
 &\ \ \ + \left. \frac{\partial\bm{F}(\bm{X},\bm{I})}{\partial\bm{I}}^{\top}
 \frac{\partial R(\bm{X},\bm{I})}{\partial\bm{X}} \right|_{\bm{X}=\bm{X}_0(\theta,\bm{I})} \cr
 &= \left.\left[\frac{\partial}{\partial\bm{X}}\left(\frac{\partial R(\bm{X},\bm{I})}{\partial\bm{I}}\right)\right]
 \bm{F}(\bm{X},\bm{I})\right|_{\bm{X}=\bm{X}_0(\theta,\bm{I})}
 + \bm{\zeta}_r(\theta,0,\bm{I}), \cr
 \label{eq. def R func def}
\end{align}
where $\partial/\partial\bm{I}$ denotes a differential operator defined as
$(\partial/\partial\bm{I})f(\bm{I})$ $=$ $[\partial f(\bm{I})/\partial I_1,$ $\ldots,$ $\partial f(\bm{I})/\partial I_m]^{\top}$ $\in\mathbb{R}^m$ for a scalar function $f(\bm{I})$,
$\frac{\partial}{\partial\bm{I}}(\frac{\partial R(\bm{X},\bm{I})}{\partial\bm{X}})$ is a matrix whose ($i,j$)-th element is given by $\frac{\partial^2 R(\bm{X},\bm{I})}{\partial X_i \partial I_j}$,
and $\frac{\partial}{\partial\bm{X}}(\frac{\partial R(\bm{X},\bm{I})}{\partial\bm{I}})$ is the transpose of $\frac{\partial}{\partial\bm{I}}(\frac{\partial R(\bm{X},\bm{I})}{\partial\bm{X}})$.
Here, the first term of the right-hand side of Eq.~(\ref{eq. def R func def}) can be written as
\begin{align}
 &\left.\left[\frac{\partial}{\partial\bm{X}}\left(\frac{\partial R(\bm{X},\bm{I})}{\partial\bm{I}}\right)\right]
 \bm{F}(\bm{X},\bm{I})\right|_{\bm{X}=\bm{X}_0(\theta,\bm{I})}
 =\left.\left[\frac{\partial}{\partial\bm{X}}\left(\frac{\partial R(\bm{X},\bm{I})}{\partial\bm{I}}\right)\right]
 \right|_{\bm{X}=\bm{X}_0(\theta,\bm{I})}
 \left.\frac{d\bm{X}_0(\omega(\bm{I})t,\bm{I})}{dt}\right|_{t=\theta/\omega(\bm{I})} \cr
 &= \omega(\bm{I}) \left.\left[\frac{\partial}{\partial\bm{X}}\left(\frac{\partial R(\bm{X},\bm{I})}{\partial\bm{I}}\right)\right]
 \right|_{\bm{X}=\bm{X}_0(\theta,\bm{I})}
 \frac{\partial\bm{X}_0(\theta,\bm{I})}{\partial\theta}
 = \omega(\bm{I}) \frac{\partial}{\partial\theta} \left. \left( \frac{\partial R(\bm{X},\bm{I})}{\partial \bm{I}} \right)
 \right|_{\bm{X}=\bm{X}_0(\theta,\bm{I})} \cr
 &= \omega(\bm{I}) \frac{\partial \bm{\xi}_r(\theta,0,\bm{I})}{\partial \theta}.
 \label{eq. def R func def - rewrite}
\end{align}
Furthermore, differentiating the right-hand side of Eq.~(\ref{eq. def R func}), we can derive
\begin{align}
 \left. \frac{\partial}{\partial\bm{I}} \left[ - \lambda(\bm{I}) R(\bm{X},\bm{I}) \right] \right|_{\bm{X}=\bm{X}_0(\theta,\bm{\bm{I}})}
 &= - \left. \left[ \frac{d\lambda(\bm{I})}{d\bm{I}} R(\bm{X},\bm{I})
 + \lambda(\bm{I}) \frac{\partial R(\bm{X},\bm{I})}{\partial\bm{I}} \right] \right|_{\bm{X}=\bm{X}_0(\theta,\bm{\bm{I}})} \cr
 &=  -  \lambda(\bm{I}) \bm{\xi}_r(\theta,0,\bm{I}),
 \label{eq. def R func def left}
\end{align}
where we used $R(\bm{X}_0(\theta,\bm{I}),\bm{I})=0$.
Thus, from Eqs.~(\ref{eq. def R func})--(\ref{eq. def R func def left}), we can obtain
\begin{align}
 \omega(\bm{I}) \frac{\partial \bm{\xi}_r(\theta,0,\bm{I})}{\partial \theta} + \bm{\zeta}_r(\theta,0,\bm{I})
 &= -  \lambda(\bm{I}) \bm{\xi}_r(\theta,0,\bm{I})
 \label{eq. R 1st ode}
\end{align}
Since Eq.~(\ref{eq. R 1st ode}) is a linear first-order ordinary differential equation for $\bm{\xi}_r(\theta,0,\bm{I})$,
this equation can be solved as follows:
\begin{align}
 \bm{\xi}_r(\theta,0,\bm{I})
 &= - \frac{1}{\omega(\bm{I})} \int_{-\infty}^{\theta} e^{\lambda(\bm{I})(\theta'-\theta)/\omega(\bm{I})} \bm{\zeta}_r(\theta',0,\bm{I}) d\theta',
\end{align}
which leads to Eqs.~(\ref{eq. xi r - xeta eta}) and~(\ref{eq. xi r - xeta eta2}).}

Using the derived Eq.~(\ref{eq. xi r - xeta eta2}),
we can estimate the order of ${\bm \xi}_{r}(\theta, 0, {\bm q}(\epsilon t))$ as
\begin{align}
 \bm{\xi}_r(\theta,0,\bm{q}(\epsilon t))
 &=\frac{1}{\lambda(\bm{I})}
	\left.\int_{0}^{\infty}
	e^{-s}
	\bm{\zeta}_r(\theta-\omega(\bm{I})s/\lambda(\bm{I}),0,\bm{I})ds
	\right|_{\bm{I}=\bm{q}(\epsilon t)} \cr
	&= \frac{1}{\lambda(\bm{I})}
	\left.\int_{0}^{\infty}
	e^{-s}
	\bm{\zeta}_r(\theta,0,\bm{I})ds
	\right|_{\bm{I}=\bm{q}(\epsilon t)}
	+ O\left(\frac{1}{\lambda(\bm{q}(\epsilon t))^2}\right)
	\cr
	&=O\left(\frac{1}{\lambda(\bm{q}(\epsilon t))}\right),
	\label{eq. order xi r}
\end{align}
where we expanded $\bm{\zeta}_r(\theta,r,\bm{I})$ in $\theta$ in the second line.
For simplicity of notation, we introduce
$\tilde{\bm{\xi}}_r(\theta,\bm{I})$ as follows:
\begin{eqnarray}
 \tilde{\bm{\xi}}_r(\theta,\bm{I})=
	\lambda(\bm{I})\bm{\xi}_r(\theta,0,\bm{I})
	=\int_{0}^{\infty}
	e^{-s}\bm{\zeta}_r(\theta-\omega(\bm{I})s
	/\lambda(\bm{I}),0,\bm{I})ds.
	\label{eq. tilde xi}
\end{eqnarray}
Note that $\tilde{\bm{\xi}}_r(\theta,\bm{I})$ is of the order $O(1)$.

To evaluate the order of $r(t)$,
we approximate the solution to Eq. (\ref{eq. r4}) describing the oscillator amplitude
in a small neighborhood of $t=t'$.
We introduce a small parameter
$\tilde{\epsilon}:=\epsilon/\lambda(\bm{q}(\epsilon t'))$,
which is  sufficiently small ($\ll1$)
by the assumption that $\epsilon\ll\lambda(\bm{q}(\epsilon t))$.
Then, using the small parameters $\sigma$ and $\tilde{\epsilon}$,
we expand the solutions to Eqs.~(\ref{eq. theta4})
and (\ref{eq. r4}) as follows:
\begin{eqnarray}
 \theta(t)&=&\theta_0(t)+\sigma\theta_{\sigma,1}(t)
	+\tilde{\epsilon}\theta_{\epsilon,1}(t)+\cdots, \\
	 r(t)&=&r_0(t)+\sigma r_{\sigma,1}(t)
	+\tilde{\epsilon} r_{\epsilon,1}(t)+\cdots,
\end{eqnarray}
where $\theta_0(t)$ and $r_0(t)$
are the lowest order solutions
and $\theta_{\sigma,j}(t)$, $r_{\sigma,j}(t)$,
$\theta_{\epsilon,j}(t)$, and $r_{\epsilon,j}(t)$
are $j$th order perturbations.
The lowest order solutions are given by $\theta_0(t)=\theta(t')+\omega(\bm{q}(\epsilon t'))(t-t')$ and $r_0(t)=0$ in the neighborhood of $t=t'$.
By introducing a rescaled time $s=\Phi(t):=\int_{0}^t\lambda(\bm{q}(\epsilon t'))dt'$ (i.e., $ds=\lambda(\bm{q}(\epsilon t))dt$),
we can rewrite Eq.~(\ref{eq. r4}) as
\begin{eqnarray}
 \frac{dr}{ds}
	=-r
	+\frac{\sigma\bm{\zeta}_r(\theta,r,\bm{q}(\epsilon t))}
	{\lambda(\bm{q}(\epsilon t))}\cdot\bm{p}(t)
	+\frac{\epsilon\bm{\xi}_r(\theta,r,\bm{q}(\epsilon t))}
	{\lambda(\bm{q}(\epsilon t))}\cdot\dot{\bm{q}}(\epsilon t).
\label{eq. r app4}
\end{eqnarray}
We also  expand $\bm{q}(\epsilon t)$
around $t=t'$ ($s=\Phi(t')$)
as $\bm{q}(\epsilon t)=\bm{q}(\epsilon t')
+\epsilon\bm{q}'(\epsilon t')(t-t')+\cdots$.
Plugging $\theta(t)=\theta_0(t)+O(\sigma,\tilde{\epsilon})$,
$r(t)=r_0(t)+O(\sigma,\tilde{\epsilon})$ and
$\bm{q}(\epsilon t)=\bm{q}(\epsilon t')+O(\epsilon)$
into Eq. (\ref{eq. r app4}), we can derive
\begin{eqnarray}
 \frac{dr}{ds}
	&=&-r
	+\sigma\frac{\bm{\zeta}_r(\theta_0(t)+O(\sigma,\tilde{\epsilon}),
	0+O(\sigma,\tilde{\epsilon}),\bm{q}(\epsilon t')+O(\epsilon))}
	{\lambda(\bm{q}(\epsilon t')+O(\epsilon))}\cdot\bm{p}(t) \cr
	&& +\epsilon\frac{\bm{\xi}_r(\theta_0(t)+O(\sigma,\tilde{\epsilon}),
  0+O(\sigma,\tilde{\epsilon}),\bm{q}(\epsilon t')+O(\epsilon))}
	{\lambda(\bm{q}(\epsilon t')+O(\epsilon))}\cdot\dot{\bm{q}}(\epsilon
	t) \cr
	&=&-r
	+\sigma(1+O(\epsilon))
	\frac{\bm{\zeta}_r(\theta_0(t)+O(\sigma,\tilde{\epsilon}),
	0+O(\sigma,\tilde{\epsilon}),\bm{q}(\epsilon t')+O(\epsilon))}
	{\lambda(\bm{q}(\epsilon t'))}\cdot\bm{p}(t) \cr
	&& +\epsilon(1+O(\epsilon))
	\frac{\bm{\xi}_r(\theta_0(t)+O(\sigma,\tilde{\epsilon}),
  0+O(\sigma,\tilde{\epsilon}),\bm{q}(\epsilon t')+O(\epsilon))}
	{\lambda(\bm{q}(\epsilon
	t'))}\cdot\dot{\bm{q}}(\epsilon t) \cr
	&=&-r
	+\sigma(1+O(\epsilon))\frac{\bm{\zeta}_r(\theta_0(t),
	0,\bm{q}(\epsilon t')+O(\epsilon))}
	{\lambda(\bm{q}(\epsilon t'))}\cdot\bm{p}(t) \cr
	&& +\epsilon(1+O(\epsilon))\frac{\bm{\xi}_r(\theta_0(t),
  0,\bm{q}(\epsilon t')+O(\epsilon))}
	{\lambda(\bm{q}(\epsilon t'))}\cdot\dot{\bm{q}}(\epsilon t)
	+O(\sigma^2,\sigma\tilde{\epsilon},\tilde{\epsilon}^2).
\end{eqnarray}
Substituting Eq. (\ref{eq. tilde xi}) into the above equation, we obtain
\begin{eqnarray}
\frac{dr}{ds}
	&=&-r
	+\sigma(1+O(\epsilon))\frac{\bm{\zeta}_r(\theta_0(t),
	0,\bm{q}(\epsilon t')+O(\epsilon))}
	{\lambda(\bm{q}(\epsilon t'))}\cdot\bm{p}(t) \cr
	&& +\epsilon(1+O(\epsilon))
	\frac{\tilde{\bm{\xi}}_r(\theta_0(t),
  \bm{q}(\epsilon t')+O(\epsilon))}
	{\lambda(\bm{q}(\epsilon t'))\lambda(\bm{q}(\epsilon
	t')+O(\epsilon))}\cdot\dot{\bm{q}}(\epsilon t)
	+O(\sigma^2,\sigma\tilde{\epsilon},\tilde{\epsilon}^2) \cr
	&=&-r
	+\sigma(1+O(\epsilon))\frac{\bm{\zeta}_r(\theta_0(t),
	0,\bm{q}(\epsilon t'))+O(\epsilon)}
	{\lambda(\bm{q}(\epsilon t'))}\cdot\bm{p}(t) \cr
	&& +\epsilon(1+O(\epsilon))\frac{\tilde{\bm{\xi}}_r(\theta_0(t),
  \bm{q}(\epsilon t'))+O(\epsilon)}
	{\lambda(\bm{q}(\epsilon t'))^2}\cdot\dot{\bm{q}}(\epsilon t)
	+O(\sigma^2,\sigma\tilde{\epsilon},\tilde{\epsilon}^2) \cr
	&=& -r
	+\sigma\frac{\bm{\zeta}_r\big(\theta(t')
	+\omega(\bm{q}(\epsilon t'))(t-t'),0,\bm{q}(\epsilon t')\big)}
	{\lambda(\bm{q}(\epsilon t'))}\cdot\bm{p}(t) \cr
	&& +\epsilon\frac{\tilde{\bm{\xi}}_r\big(\theta(t')
	+\omega(\bm{q}(\epsilon t'))(t-t'),\bm{q}(\epsilon t')\big)}
	{\lambda(\bm{q}(\epsilon t'))^2}\cdot\dot{\bm{q}}(\epsilon t)
	+O(\sigma^2,\sigma\tilde{\epsilon},\tilde{\epsilon}^2).
\label{eq. r app7}
\end{eqnarray}
By integrating Eq.~(\ref{eq. r app7}), we can estimate the order of $r(t')$ as
\begin{align}
 r(t')&=\frac{\sigma}{\lambda(\bm{q}(\epsilon t'))}
	\int_{-\infty}^{\Phi(t')}e^{s-\Phi(t')}\bm{\zeta}_r
	\big(\theta(t')+\omega(\bm{q}(\epsilon t'))(t-t'),
	0,\bm{q}(\epsilon t')\big)
	\cdot\bm{p}(t)\big|_{t=\Phi^{-1}(s)}ds \cr
	& +\frac{\epsilon}{\lambda(\bm{q}(\epsilon t'))^2}
	\int_{-\infty}^{\Phi(t')}e^{s-\Phi(t')}\tilde{\bm{\xi}}_r
	\big(\theta(t')+\omega(\bm{q}(\epsilon t'))(t-t'),
	\bm{q}(\epsilon t')\big)
	\cdot\dot{\bm{q}}(\epsilon t)\big|_{t=\Phi^{-1}(s)}ds
	+O(\sigma^2,\sigma\tilde{\epsilon},\tilde{\epsilon}^2) \cr
	&=
	O\left(\frac{\sigma}{\lambda(\bm{q}(\epsilon t'))},
			 \frac{\epsilon}{\lambda(\bm{q}(\epsilon t'))^2}\right).
	\label{eq. prder r}
\end{align}
Now, by  expanding Eq.~(\ref{eq. theta3}) in $r$, 
we can obtain
\begin{eqnarray}
 \dot{\theta}
	&=&\omega(\bm{q}(\epsilon t))+\sigma
	\bm{\zeta}_{\theta}(\theta,0,\bm{q}(\epsilon t))
	\cdot\bm{p}(t)
	+\epsilon \bm{\xi}_{\theta}(\theta,0,\bm{q}(\epsilon t))
	\cdot\dot{\bm{q}}(\epsilon t)
	\nonumber \\
	&&+\sigma
	 r\frac{\partial\bm{\zeta}_{\theta}(\theta,0,\bm{q}(\epsilon t))}
	 {\partial r}\cdot\bm{p}(t)
	+\epsilon r\frac{\partial\bm{\xi}_{\theta}(\theta,0,\bm{q}(\epsilon t))}
	{\partial r}\cdot\dot{\bm{q}}(\epsilon t)
	+O(r^{2}).
	\label{eq. phase eq1}
\end{eqnarray}
Substituting Eq.~(\ref{eq. prder r}) into Eq.~(\ref{eq. phase eq1}) and neglecting higher order terms in $r$,
we can derive the generalized phase equation (\EQPhaseEq) in the main article,
{
\begin{eqnarray}
 \dot{\theta} &=& \omega(\bm{q}(\epsilon t)) +\sigma \bm{\zeta}_{\theta}(\theta,0,\bm{q}(\epsilon t)) \cdot\bm{p}(t)
	+ O\left(\frac{\sigma^2}{\lambda(\bm{q}(\epsilon t))}, \frac{\sigma\epsilon}{\lambda(\bm{q}(\epsilon t))^2}\right)\cr
 && + \epsilon \bm{\xi}_{\theta}(\theta,0,\bm{q}(\epsilon t)) \cdot \dot{\bm{q}}(\epsilon t)
 + O\left(\frac{\sigma\epsilon}{\lambda(\bm{q}(\epsilon t))}, \frac{\epsilon^2}{\lambda(\bm{q}(\epsilon t))^2}\right).
	\label{eq. phase eq2}
\end{eqnarray}
}

Equation~(\ref{eq. phase eq2}) reveals that our phase equation  well approximates
the exact phase dynamics under the conditions that{
\begin{align}
 \frac{\sigma^2}{\lambda(\bm{q}(\epsilon t))} \ll \sigma,
 \,\,\,
 \frac{\sigma\epsilon}{\lambda(\bm{q}(\epsilon t))^2}  \ll \sigma,
 \,\,\,
 \frac{\sigma\epsilon}{\lambda(\bm{q}(\epsilon t))}  \ll \epsilon,
 \,\,\,{\rm and}\,\,\,
 \frac{\epsilon^2}{\lambda(\bm{q}(\epsilon t))^2} \ll \epsilon.
\end{align}
Here, we compared the first two error terms $\frac{\sigma^2}{\lambda(\bm{q}(\epsilon t))}$
and $\frac{\sigma\epsilon}{\lambda(\bm{q}(\epsilon t))^2}$ with $\sigma$,
and the last two $\frac{\sigma\epsilon}{\lambda(\bm{q}(\epsilon t))}$
and $\frac{\epsilon^2}{\lambda(\bm{q}(\epsilon t))^2}$ with $\epsilon$,
because the first and last two error terms arose when we expanded the second term
$\sigma \bm{\zeta}_{\theta}(\theta,r,\bm{q}(\epsilon t)) \cdot \bm{p}(t)$ ($=O(\sigma)$)
and the third term
$\epsilon \bm{\xi}_{\theta}(\theta,r,\bm{q}(\epsilon t)) \cdot \dot{\bm{q}}(\epsilon t)$ ($=O(\epsilon)$)
of Eq.~(\ref{eq. theta4}) in $r$, respectively.
because the first two terms arise from the expansion of the second term
$\sigma \bm{\zeta}_{\theta}(\theta,r,\bm{q}(\epsilon t)) \cdot \bm{p}(t)$ ($=O(\sigma)$) of Eq.~(\ref{eq. theta4}) in $r$,
and the last two terms arise from the third term
$\epsilon \bm{\xi}_{\theta}(\theta,r,\bm{q}(\epsilon t)) \cdot \dot{\bm{q}}(\epsilon t)$ ($=O(\epsilon)$), respectively.}
These conditions are satisfied when 
{
\begin{eqnarray}
	\frac{\epsilon}{\lambda(\bm{q}(\epsilon t))^2} \ll 1
	\,\,\,{\rm and}\,\,\,
	\frac{\sigma}{\lambda(\bm{q}(\epsilon t))} \ll 1,
	\label{eq. error order pe}
\end{eqnarray}
}namely, when (i) the timescale of the slowly varying component $\bm{q}(\epsilon t)$ is much larger than the relaxation time of
perturbed orbits to $C$, and (ii) the remaining fluctuations $\sigma\bm{p}(t)$ is sufficiently weak, as we assumed in the main article.

For limit-cycle oscillators
with higher-dimensional state variables ($n\ge3$),
we can also derive a phase equation
corresponding to Eq.~(\ref{eq. phase eq2}).
In higher-dimensional cases,
the system of Eq.~(\ref{eq. model0}) has $n$ ($\ge3$) Floquet exponents.
Let $\lambda_j(\bm{I})$ denote the absolute value of the $j$-th largest Floquet exponent of the oscillator for a given constant $\bm{I}$
($\lambda_1(\bm{I})=0 > \lambda_2(\bm{I}) \ge \cdots \ge \lambda_n(\bm{I})$).
In these exponents, the second largest exponent $\lambda_2(\bm{I})$ dominates the relaxation time of perturbed orbits.
Thus, using the absolute value of the second largest Floquet exponent $\lambda_2(\bm{q}(\epsilon t))$
instead of $\lambda(\bm{q}(\epsilon t))$,
we can obtain the same results as Eq.~(\ref{eq. phase eq2});
that is, we can obtain the following phase equation also
for the higher-dimensional cases ($n\ge3$):
\begin{eqnarray}
 \dot{\theta}
	&=&\omega(\bm{q}(\epsilon t))+\sigma
	\bm{\zeta}_{\theta}(\theta,0,\bm{q}(\epsilon t))
	\cdot\bm{p}(t)
	+\epsilon \bm{\xi}_{\theta}(\theta,0,\bm{q}(\epsilon t))
	\cdot\dot{\bm{q}}(\epsilon t)
	\nonumber \\
	&& +O\left(\frac{\epsilon^2}{\lambda_2(\bm{q}(\epsilon t))^2},
	 \frac{\sigma\epsilon}{\lambda_2(\bm{q}(\epsilon t))},
	 \frac{\sigma^2}{\lambda_2(\bm{q}(\epsilon t))}\right).
	\label{eq. phase eq3}
\end{eqnarray}

\section{Relations among different sensitivity functions}

This section gives a derivation of
Eqs.~(\EQPropZZetaXiA)--(\EQPropOmegZeta) in the main article.
These relations are essentially important in
understanding the properties of the sensitivity functions
and in developing methods to calculate and estimate the sensitivity functions.
In this section, for simplicity of notation,
the sensitivity functions are denoted by
$\bm{\zeta}(\theta,\bm{I})$
and $\bm{\xi}(\theta,\bm{I})$
as in the main article.

\subsection{Derivation of Eq.~(\EQPropZZetaXiA) in the main article}

As we shown in Eq.~(\EQPropZZetaXiA) in the main article,
the sensitivity function $\bm{\xi}(\theta,\bm{I})$ can be written as
\begin{eqnarray}
 \bm{\xi}(\theta,\bm{I})
	=-\frac{\partial\bm{X}_0(\theta,\bm{I})}
	{\partial \bm{I}}^{\top}
	\bm{Z}(\theta,\bm{I}).
	\label{eq. xi - Z}
\end{eqnarray}
This equation relates the change in the shape of the limit-cycle orbit
$\bm{X}_{0}(\theta, {\bm I})$ and the phase sensitivity function
${\bm Z}(\theta,\bm{I})$ to the sensitivity function ${\bm \xi}(\theta, {\bm I})$.
{From the definition of $\Theta(\bm{X},\bm{I})$,
\begin{eqnarray}
 \Theta(\bm{X}_0(\theta,\bm{I}),\bm{I})=\theta
	\label{eq. trib rel phase}
\end{eqnarray}
holds. 
By differentiating Eq.~(\ref{eq. trib rel phase}) with respect to $\bm{I}$, we can obtain
\begin{align}
 \frac{\partial}{\partial\bm{I}} \Theta(\bm{X}_0(\theta,\bm{I}),\bm{I}) 
 &= \frac{\partial\bm{X}_0(\theta,\bm{I})}{\partial \bm{I}}^{\top}
 \left.\frac{\partial\Theta(\bm{X},\bm{I})}{\partial\bm{X}}\right|_{\bm{X}=\bm{X}_0(\theta,\bm{I})}
 + \left.\frac{\partial\Theta(\bm{X},\bm{I})}{\partial\bm{I}}\right|_{\bm{X} = \bm{X}_0(\theta,\bm{I})} \cr
 &= \frac{\partial\bm{X}_0(\theta,\bm{I})}{\partial \bm{I}}^{\top} \bm{Z}(\theta,\bm{I})
 + \bm{\xi}(\theta,\bm{I})
 = 0,
\end{align}
which leads to Eq.~(\ref{eq. xi - Z}).}

\subsection{Derivation of Eqs.~(\EQPropZZetaXiB) and (\EQPropOmegZeta) in the main article}

As we shown in Eqs.~(\EQPropZZetaXiB) and (\EQPropOmegZeta) in the main article,
the sensitivity functions $\bm{\zeta}(\theta,\bm{I})$ and
$\bm{\xi}(\theta,\bm{I})$ are mutually related as follows:
\begin{eqnarray}
 \bm{\xi}(\theta,\bm{I})&=&\bm{\xi}(\theta_0,\bm{I})
	-\frac{1}{\omega(\bm{I})}
	\int_{\theta_0}^{\theta} [\bm{\zeta}(\phi,\bm{I})
	-\bar{\bm{\zeta}}(\bm{I})]d\phi,
	\label{eq. YZ rel1}
\end{eqnarray}
\begin{eqnarray}
 \bm{\zeta}(\theta,\bm{I})&=& \bar{\bm{\zeta}}(\bm{I})
	-\omega(\bm{I})\frac{\partial \bm{\xi}(\theta,\bm{I})}{\partial\theta},
	\label{eq. YZ rel2}
\end{eqnarray}
{and
\begin{eqnarray}
  \bar{\bm{\zeta}}(\bm{I})
	 := \frac{1}{2\pi}\int_0^{2\pi}\bm{\zeta}(\theta,\bm{I})d\theta
	 = \frac{d\omega(\bm{I})}{d\bm{I}},
	 \label{eq. rel omega zeta}
\end{eqnarray}
}where $\theta_0 \in [0,2\pi)$ is an arbitrary phase and
$\bar{\bm{\zeta}}(\bm{I})$ is the average of $\bm{\zeta}(\theta,\bm{I})$ with respect to $\theta$ and is a function of $\bm{I}$.
{Equation~(\ref{eq. YZ rel1}) (or~(\ref{eq. YZ rel2})) represents the sensitivity function ${\bm \xi}(\theta, {\bm I})$
characterizing the phase response caused by a small constant shift in ${\bm I}$
as an integral of the phase response to the instantaneous change in ${\bm I}$ at each $\theta$,
and Eq.~(\ref{eq. rel omega zeta}) relates the change in the frequency $\omega({\bm I})$ of the limit-cycle orbit
to the average of the sensitivity function ${\bm \zeta}(\theta,\bm{I})$,
i.e., the net phase shift caused by the a small constant shift in ${\bm I}$ during one period of oscillation.}
Using these relations, we can obtain the sensitivity function
${\bm \xi}(\theta, {\bm I})$ for each ${\bm I}$. Namely,
we can calculate the sensitivity function ${\bm \zeta}(\theta, {\bm I})$,
e.g.,  by using the adjoint method,
and then integrate ${\bm \zeta}(\theta, {\bm I})$ with respect to $\theta$
to obtain the sensitivity function ${\bm \xi}(\theta, {\bm I})$.

{Since we can straightforwardly derive Eq.~(\ref{eq. YZ rel1}) by integrating Eq.~(\ref{eq. YZ rel2}) with respect to $\theta$,
we only describe derivations of Eq.~(\ref{eq. YZ rel2})
and  Eq.~(\ref{eq. rel omega zeta}).
From the definition of $\Theta(\bm{X},\bm{I})$,
\begin{align}
 \frac{\partial\Theta(\bm{X},\bm{I})}{\partial\bm{X}} \cdot \bm{F}(\bm{X},\bm{I}) &= \omega(\bm{I})
 \label{eq. def phase func}
\end{align}
holds. By differentiating Eq.~(\ref{eq. def phase func}) with respect to $\bm{I}$ and plugging in $\bm{X}=\bm{X}_0(\theta,\bm{I})$,
we can obtain
	\begin{align}
\left. \frac{\partial}{\partial\bm{I}} \left[ \frac{\partial \Theta(\bm{X},\bm{I})}{\partial\bm{X}} \cdot \bm{F}(\bm{X},\bm{I}) \right] \right|_{\bm{X} = \bm{X}_0(\theta,\bm{I})}
 &= \left.\left[\frac{\partial}{\partial\bm{I}}\left(\frac{\partial \Theta(\bm{X},\bm{I})}{\partial\bm{X}}\right)\right]^{\top}
 \bm{F}(\bm{X},\bm{I}) \right|_{\bm{X}=\bm{X}_0(\theta,\bm{I})}
 \cr
 &\ \ \ + \left. \frac{\partial\bm{F}(\bm{X},\bm{I})}{\partial\bm{I}}^{\top}
 \frac{\partial \Theta(\bm{X},\bm{I})}{\partial\bm{X}} \right|_{\bm{X}=\bm{X}_0(\theta,\bm{I})} \cr
 &= \left.\left[\frac{\partial}{\partial\bm{X}}\left(\frac{\partial \Theta(\bm{X},\bm{I})}{\partial\bm{I}}\right)\right]
 \bm{F}(\bm{X},\bm{I})\right|_{\bm{X}=\bm{X}_0(\theta,\bm{I})}
 + \bm{\zeta}(\theta,\bm{I}) \cr
 &= \frac{d \omega(\bm{I})}{d\bm{I}},
 \label{eq. def phase func def}
	\end{align}
where $\frac{\partial}{\partial\bm{I}}(\frac{\partial \Theta(\bm{X},\bm{I})}{\partial\bm{X}})$ is a matrix whose ($i,j$)-th element is given by
$\frac{\partial^2 \Theta(\bm{X},\bm{I})}{\partial X_i \partial I_j}$,
and $\frac{\partial}{\partial\bm{X}}(\frac{\partial \Theta(\bm{X},\bm{I})}{\partial\bm{I}})$
is the transpose of $\frac{\partial}{\partial\bm{I}}(\frac{\partial \Theta(\bm{X},\bm{I})}{\partial\bm{X}})$.
Here, the first term of the third line in Eq.~(\ref{eq. def phase func def}) can be written as
\begin{align}
 &\left.\left[\frac{\partial}{\partial\bm{X}}\left(\frac{\partial \Theta(\bm{X},\bm{I})}{\partial\bm{I}}\right)\right]
 \bm{F}(\bm{X},\bm{I})\right|_{\bm{X}=\bm{X}_0(\theta,\bm{I})}
 =\left.\left[\frac{\partial}{\partial\bm{X}}\left(\frac{\partial \Theta(\bm{X},\bm{I})}{\partial\bm{I}}\right)\right]
 \right|_{\bm{X}=\bm{X}_0(\theta,\bm{I})}
 \left.\frac{d\bm{X}_0(\omega(\bm{I})t,\bm{I})}{dt}\right|_{t=\theta/\omega(\bm{I})} \cr
 &= \omega(\bm{I}) \left.\left[\frac{\partial}{\partial\bm{X}}\left(\frac{\partial \Theta(\bm{X},\bm{I})}{\partial\bm{I}}\right)\right]
 \right|_{\bm{X}=\bm{X}_0(\theta,\bm{I})}
 \frac{\partial\bm{X}_0(\theta,\bm{I})}{\partial\theta}
 = \omega(\bm{I}) \frac{\partial}{\partial\theta} \left. \left( \frac{\partial \Theta(\bm{X},\bm{I})}{\partial \bm{I}} \right)
 \right|_{\bm{X}=\bm{X}_0(\theta,\bm{I})} \cr
 &= \omega(\bm{I}) \frac{\partial \bm{\xi}(\theta,\bm{I})}{\partial \theta}.
 \label{eq. def phase func def - rewrite}
\end{align}
Then, from Eqs.~(\ref{eq. def phase func def}) and~(\ref{eq. def phase func def - rewrite}),
we can derive Eq.~(\ref{eq. YZ rel2}) and Eq.~(\ref{eq. rel omega zeta}) as
\begin{align}
 \frac{d\omega(\bm{I})}{d\bm{I}}
 &= \frac{1}{2\pi}\int_0^{2\pi} \left[ \omega(\bm{I}) \frac{\partial \bm{\xi}(\theta,\bm{I})}{\partial \theta} + \bm{\zeta}(\theta,\bm{I}) \right] d\theta
 = \frac{1}{2\pi}\int_0^{2\pi} \bm{\zeta}(\theta,\bm{I}) d\theta,
\end{align}
where the first term in the integral vanishes due to $2\pi$-periodicity of $\bm{\xi}(\theta, {\bm I})$.}

{
\section{Relation between the conventional and generalized phase equations}

Here we compare the generalized phase equation with the
conventional phase equation using the near-identity transformation.
As stated in the main article, the conventional phase equation can be written as
\begin{align}
 \dot{\tilde{\theta}} &= \omega(\bm{q}_c) + \sigma_c \bm{\zeta}(\tilde{\theta},\bm{q}_c) \cdot \bm{p}_c(t) + O(\sigma_c^2), \label{eq. analy comp phase eq}
\end{align}
where  $\bm{q}_c \in A$  is a constant, 
$\bm{p}_c(t)$ is an external input defined as $\sigma_c\bm{p}_c(t)=\bm{I}(t)-\bm{q}_c$, and $\sigma_c$ is a parameter representing the intensity of the external input.
We decompose the external input $\bm{p}_c(t)$ into two terms, $\bm{p}_1(t)$ and $\bm{p}_2(t)$, as 
\begin{align}
 \bm{p}_c(t) &= \bm{p}_1(t) + \bm{p}_2(t),
\end{align}
and introduce a slightly deformed phase $\phi(t)$ as

\begin{align}
 \phi(t) &= \tilde{\theta}(t) + \sigma_c \bm{\xi}(\tilde{\theta}(t),\bm{q}_c) \cdot \bm{p}_1(t). \label{eq. analy comp nit}
\end{align}

By applying the above near-identity transformation  to the phase equation (\ref{eq. analy comp phase eq}),
we can derive the following phase equation for $\phi(t)$:
\begin{align}
 \dot{\phi} &= \omega(\bm{q}_c) + \sigma_c \left.\frac{d\omega(\bm{I})}{d\bm{I}}\right|_{\bm{I}=\bm{q}_c} \cdot \bm{p}_1(t)
 + \sigma_c \bm{\zeta}(\phi,\bm{q}_c) \cdot \bm{p}_2(t) + \sigma_c \bm{\xi}(\phi,\bm{q}_c) \cdot \dot{\bm{p}}_1(t) + O(\sigma_c^2).
 \label{eq. analy comp phase eq 2}
\end{align}
Without loss of generality, we can regard the input terms $\sigma_c \bm{p}_1(t)$ and $\sigma_c \bm{p}_2(t)$ in Eq.~(\ref{eq. analy comp
phase eq 2}) as the slowly varying part $\bm{q}(\epsilon t)$ and the weak fluctuations $\sigma \bm{p}(t)$ in the main article, 
because we can choose the decomposition of $\bm{p}_c(t)$ arbitrarily.
Then, Eq.~(\ref{eq. analy comp phase eq 2}) can be considered an approximation to the generalized phase equation (2) in the main article.
In other words, the first term of Eq.~(\ref{eq. analy comp phase eq 2}) represents the first-order (linear)  approximation in $\bm{q}$ around $\bm{q}=\bm{q}_c$
to the first term of the generalized phase equation,
while the second and third terms of Eq.~(\ref{eq. analy comp phase eq 2}) are zeroth-order (constant)  approximations in $\bm{q}$ around $\bm{q}=\bm{q}_c$
to the second  and third terms of the generalized phase equation.

In this sense, the generalized phase equation (2) in the main article can be considered a nonlinear generalization of the conventional phase equation (\ref{eq. analy comp phase eq 2}).
For the modified Stuart-Landau oscillator defined in the main article,
the frequency $\omega(I)$ and the sensitivity functions $\zeta(\theta,I)$ and $\xi(\theta,I)$ are explicitly given by
\begin{align}
 \omega(I) &= e^{2I} = 1 + 2I + 2I^2 + \frac{4I^3}{3} + \cdots, \\
 \zeta(\theta,I) &= 2e^{2I} - e^I\cos\theta = 1 - \cos\theta + (4 - \cos\theta) I + \left(4 - \frac{\cos\theta}{2}\right)I^2 + \cdots, \\
 \xi(\theta,I) &= e^{-I}\sin\theta = \sin\theta - (\sin\theta) I + \frac{\sin\theta}{2} I^2 + \cdots.
\end{align}
When the temporal variation in the input
$I(t)$ is sufficiently small, we can truncate $\omega(I)$ at the first order, and $\zeta(\theta, I)$ and $\xi(\theta, I)$ at the zeroth order, which is equivalent to using the conventional phase equation.
However, when the input $I(t)$ varies largely with time and the shape of the limit-cycle orbit is significantly deformed, the above approximation is no longer valid.
In such cases, the conventional phase equation would fail to predict the actual oscillator dynamics and the generalized phase reduction method should be used.}

{
\section{Accuracy and robustness of the generalized phase equation}

\begin{figure*}[tbp]
 \begin{center}
	\subfigure[$\lambda_0=1.000,\,\sigma=0.001$]{\includegraphics[width=70mm]{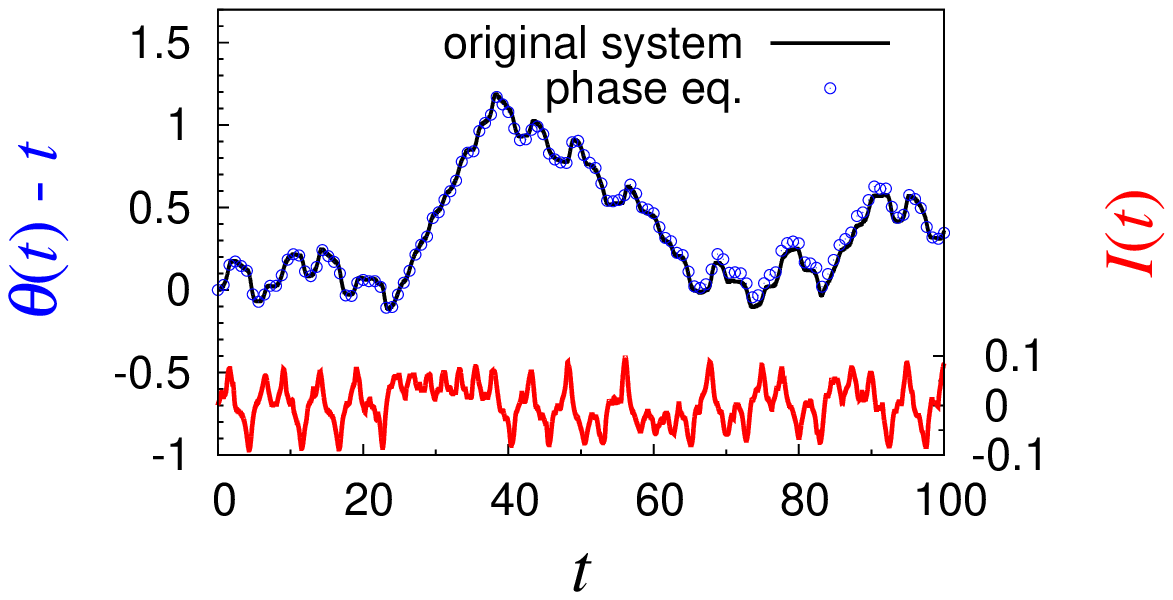}}
	\subfigure[$\lambda_0=0.215,\,\sigma=0.001$]{\includegraphics[width=70mm]{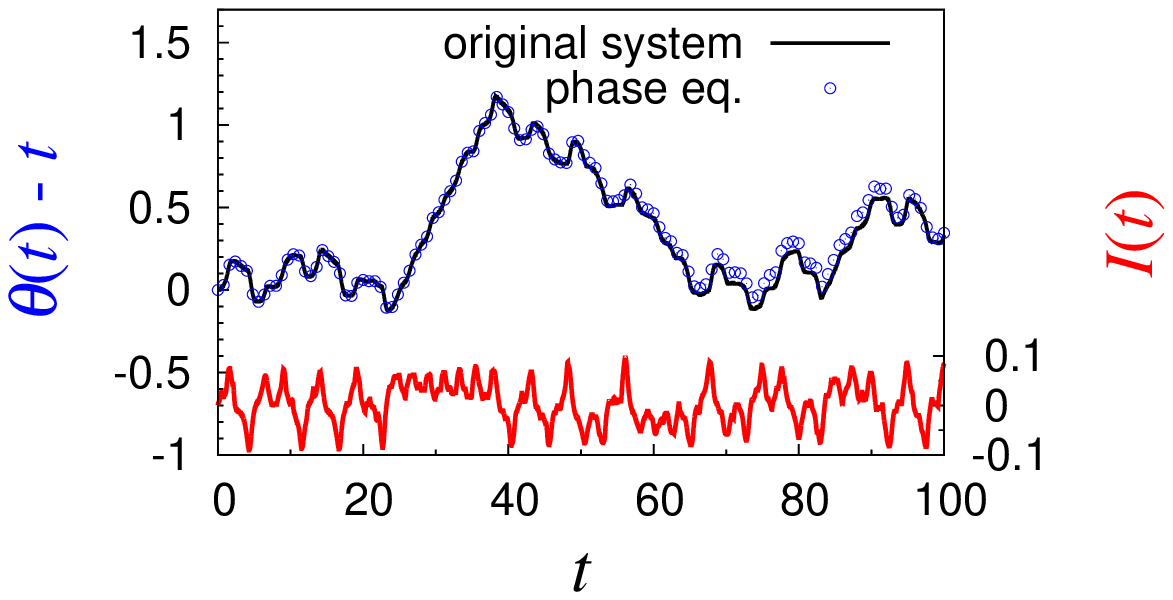}}
	\subfigure[$\lambda_0=0.046,\,\sigma=0.001$]{\includegraphics[width=70mm]{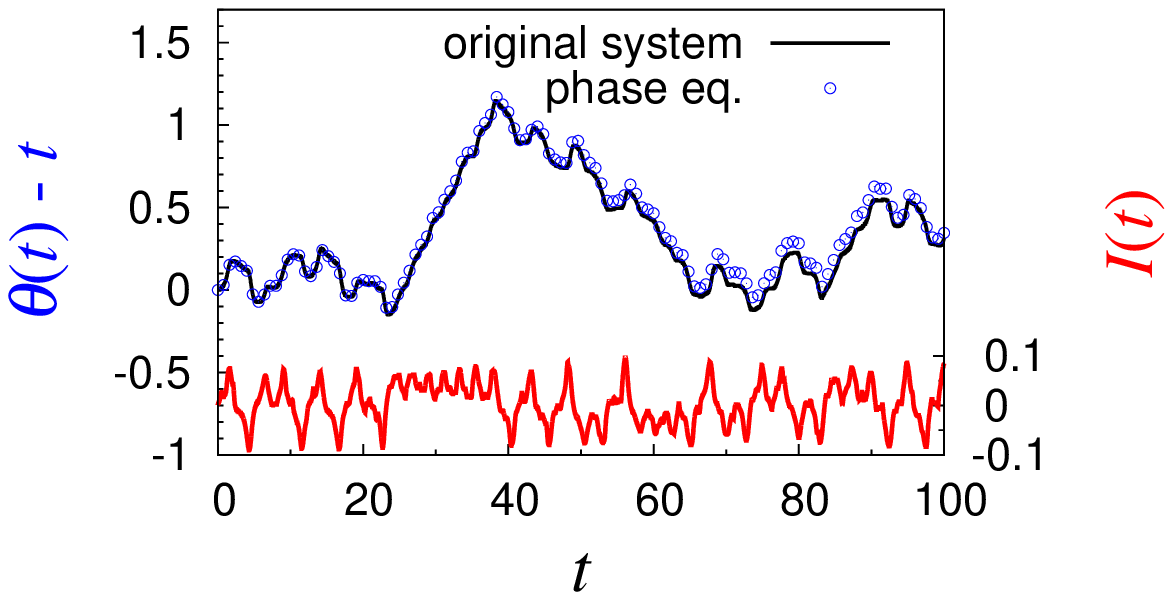}}
	\subfigure[$\lambda_0=0.010,\,\sigma=0.001$]{\includegraphics[width=70mm]{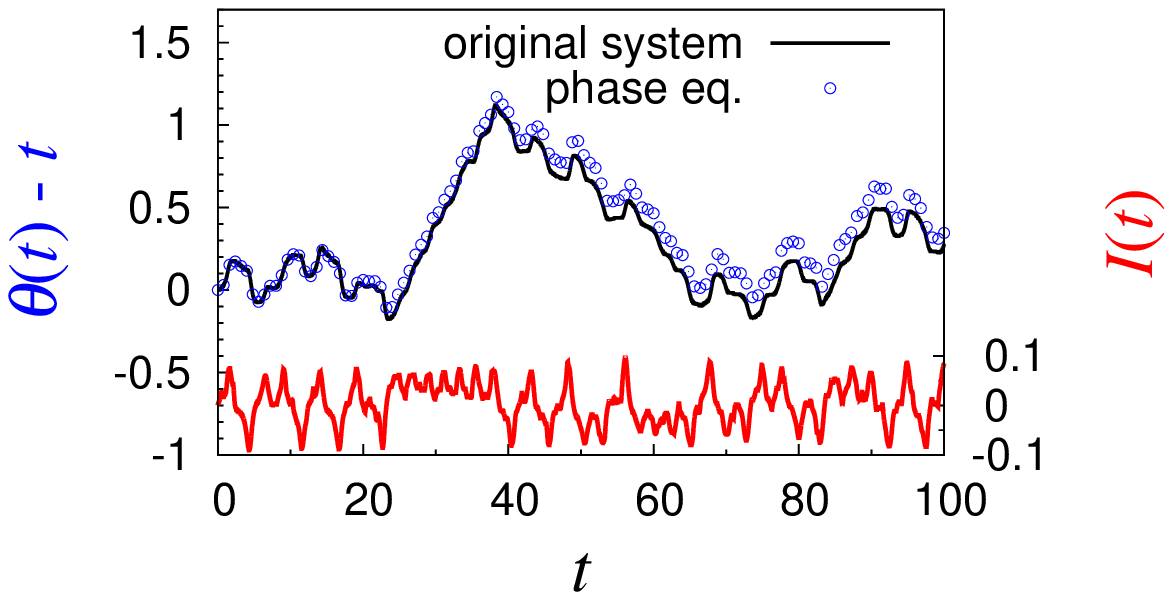}}
	\subfigure[$\lambda_0=1.000,\,\sigma=0.001$]{\includegraphics[width=70mm]{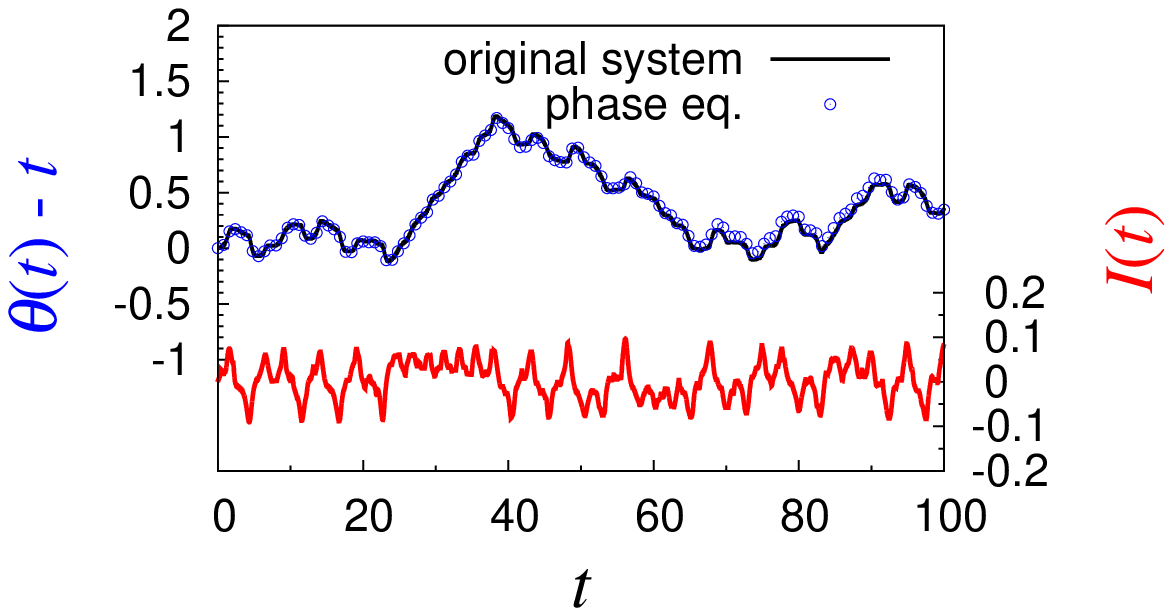}}
	\subfigure[$\lambda_0=1.000,\,\sigma=0.003$]{\includegraphics[width=70mm]{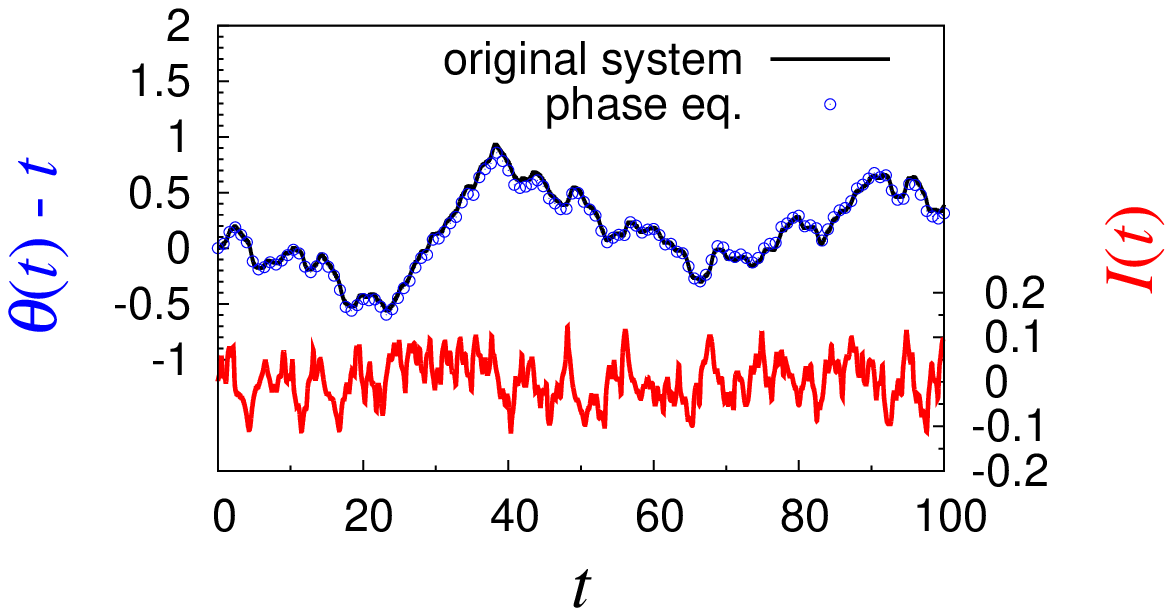}}
	\subfigure[$\lambda_0=1.000,\,\sigma=0.005$]{\includegraphics[width=70mm]{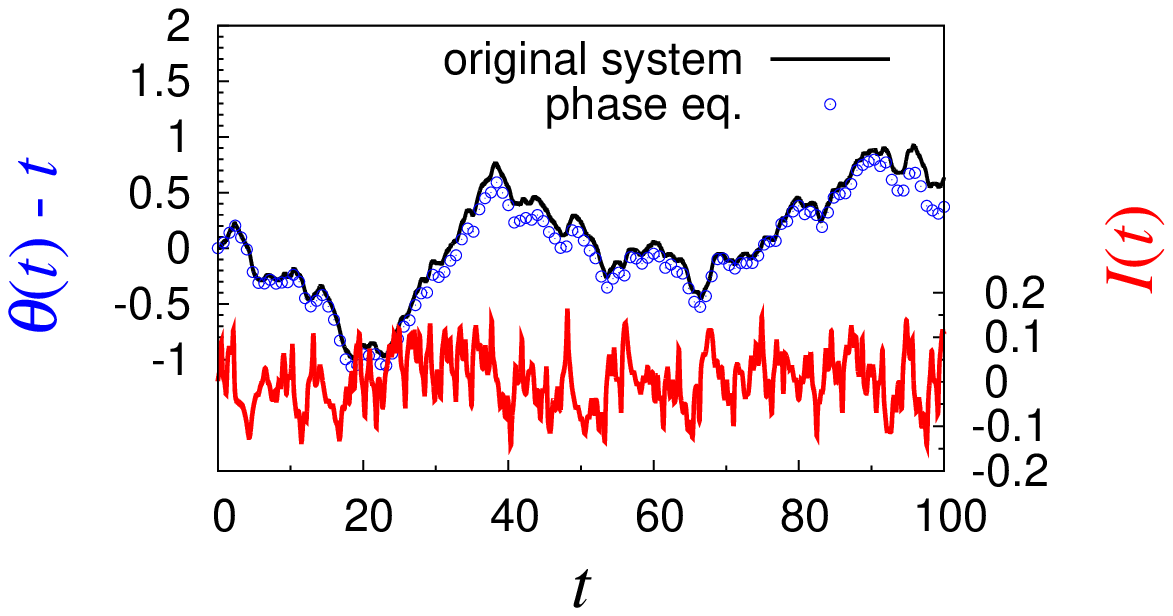}}
	\subfigure[$\lambda_0=1.000,\,\sigma=0.007$]{\includegraphics[width=70mm]{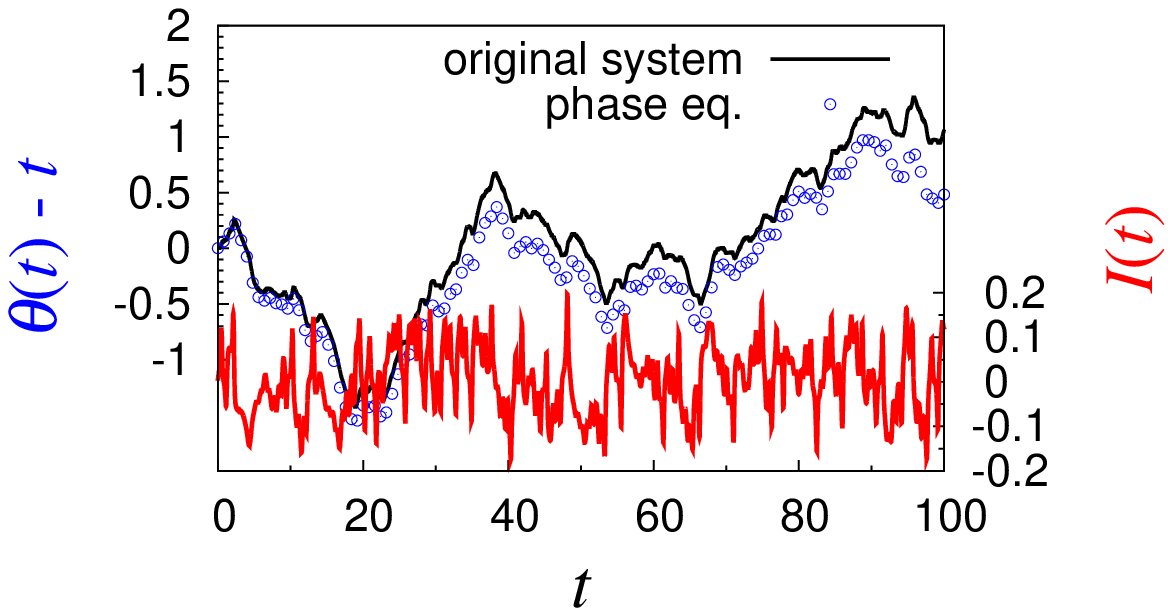}}
 \end{center}
 \caption{{(Color online)  Accuracy and robustness of the generalized phase equation.
 A modified Stuart-Landau oscillator (Eqs.~(\ref{eq. rob msl1}) and~(\ref{eq. rob msl2})) is driven
 by a periodically varying parameter $I(t)$ (red lines, Eq.~(\ref{eq. rob I})).
 The time series of the phase $\theta(t)=\Theta(\bm{X}(t),\bm{q}(\epsilon t))$ measured directly from the original system (black lines)
 and predicted by the direct numerical simulation of the generalized phase equation (blue circles) are plotted.
 In (a)--(d), $\sigma$ is fixed at $0.001$ and $\lambda_0$ is varied between $1$ and $0.01$, while
 in (e)--(h), $\lambda_0$ is fixed at $1$ and $\sigma$ is varied between $0.001$ and $0.007$.}}
 \label{fig. 1}
\end{figure*}

In the main article, we briefly demonstrated that the generalized phase equation can accurately predict the time series of the oscillation phase as compared to the conventional phase equation.
Here, we examine the accuracy and robustness of the generalized phase equation in more detail with numerical simulations. 
We use a modified Stuart-Landau oscillator defined as 
\begin{align}
 \dot{x} &= e^{2I(t)}(\lambda_0 x - y - \lambda_0 I(t)) - \lambda_0 [(x - I(t))^2 + y^2](x - I(t)), \label{eq. rob msl1} \\
 \dot{x} &= e^{2I(t)}(x + \lambda_0 y - I(t)) - \lambda_0 [(x - I(t))^2 + y^2]y, \label{eq. rob msl2}
\end{align}
whose amplitude relaxation rate can explicitly be specified by the parameter $\lambda_{0}$.
Here, $x$ and $y$ are state variables representing the oscillator state, $I(t)$ is an external input, and $\lambda_0$ is a parameter that
controls the timescale of the amplitude relaxation.
For this model, one can explicitly define the amplitude $r=\sqrt{(x - I(t))^2 + y^2}$, which decays exponentially as $\dot{r}=-2\lambda_0r$.
As stated in the main article, the small parameter $\epsilon$ represents the relative timescale of the slowly varying component $\bm{q}(\epsilon t)$
 to the amplitude relaxation time of the oscillator (which was assumed to be $O(1)$ in the main article).
Thus, by varying the parameter $\lambda_0$, we can effectively control the parameter $\epsilon$.

We applied a periodically varying parameter 
\begin{align}
 I(t) &= 0.005 L_1(0.3 t) + \sigma L_2(t)
 \label{eq. rob I}
\end{align}
to the oscillator,
where $L_1(t)$ and $L_2(t)$ are independently generated time series of the variable $x$ of the chaotic Lorenz model~\cite{SUCNS},
$\dot{x}=10(y-x)$, $\dot{y}=x(28-z)-y$, and $\dot{z}=xy-8z/3$,
and $\sigma$ is a parameter controlling the intensity of the high-frequency components in $I(t)$.
Since the parameters $\lambda_0$ and $\sigma$ play important roles in the proposed phase reduction method, we examine the accuracy and robustness of the generalized phase equation for varying values of $\lambda_{0}$ and $\sigma$.  

Figure \ref{fig. 1} shows the results of  numerical simulations, where one of the parameters is kept fixed and the other is varied.
In Figs.~\ref{fig. 1} (a)--(d),  $\sigma$ is fixed and $\lambda_0$ is varied.
The accuracy of the proposed phase reduction method is deteriorated as $\lambda_0$ is decreased.
In this case, when $\lambda_0 > 0.01$, the generalized phase equation can predict the temporal evolution of the actual phase of the oscillator.
Similarly, when $\lambda_0$ is fixed and $\sigma$ is varied (Figs.~\ref{fig. 1} (e)--(h)),
the  accuracy of the proposed method becomes worse as $\sigma$ is increased.
In this case, when $\sigma < 0.007$, the generalized phase equation can predict the temporal evolution of the actual phase.}

{
\section{Phase locking of the Morris-Lecar model driven by strong periodic forcing}

In the main article, we analyzed the phase locking of a modified Stuart-Landau oscillator to periodic forcing
and demonstrated the usefulness of the proposed phase reduction method.
In this section, we further analyze another type of limit-cycle oscillator,
i.e., the Morris-Lecar model~\cite{MFN}, which describes periodic firing of a neuron.
We theoretically analyze the phase locking dynamics of the Morris-Lecar model to  periodic external forcing
and compare the theoretical predictions with direct numerical simulations.

\subsection{The Morris-Lecar model}

The Morris-Lecar model~\cite{MFN} of a periodically firing neuron has a two-dimensional state variable $\bm{X}(t)=[V(t),w(t)]^{\top}$.
The vector field $\bm{F}(\bm{X},I)=[F_1(V,w,I),F_2(V,w,I)]^{\top}$ is given by
\begin{eqnarray}
 C_{\rm m} F_1&=&g_{\rm L}(V_L-V) + g_{\rm K}w(V_K-V) + g_{\rm Ca}m_{\infty}(V_{\rm Ca}-V)+I, \\
 F_2&=&\lambda_w(w_{\infty}-w),
\end{eqnarray}
where $m_{\infty}(V) = 0.5\{1+\tanh[(V-V_1)/V_2]\}$ and $w_{\infty}(V) = 0.5\{1+\tanh[(V-V_3)/V_4]\}$ are the conductance functions,
$I$ is the parameter to which the forcing is applied,
and $V_{\rm K}$, $V_{\rm L}$, $V_{\rm Ca}$, $g_{\rm K}$, $g_{\rm L}$,
$g_{\rm Ca}$, $C$, $V_1$, $V_2$, $V_3$, $V_4$, and $\lambda_w$ are constant parameters.
This model exhibits stable limit-cycle oscillations when the parameter values are chosen appropriately.


\subsection{Smooth oscillations}

\begin{figure*}[tbp]
 \begin{center}
	\includegraphics[width=160mm]{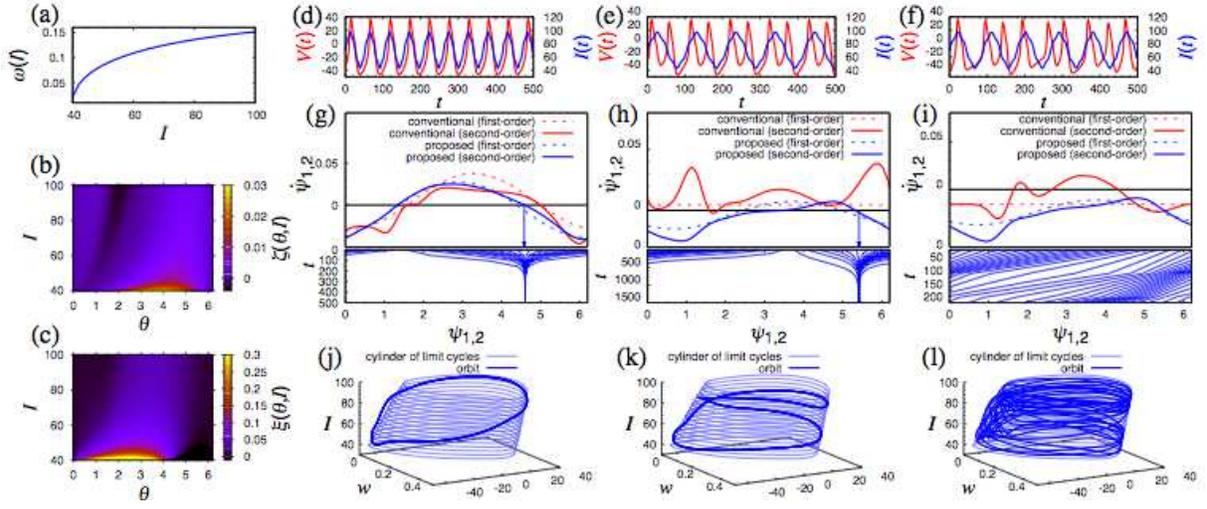}
 \end{center}
 \caption{{(Color online)
 Phase locking of the Morris-Lecar model exhibiting smooth oscillations.
 Three sets of periodically varying parameters, $I^{(j)}(t)$~: $q^{(j)}(\epsilon t)=70+25\sin(\omega_I^{(j)}t)$
 and $\sigma p^{(j)}(t)=2\sin(5\omega_I^{(j)}t)$ with $\omega_I^{(1,2,3)}$ = 0.12, 0.06, 0.07 are used, which lead to  $1:1$ or $1:2$ phase locking to $q(\epsilon t)$;
 1 : 1 phase locking to
 $I^{(1)}(t)$ [(d), (g), and (j)],
 1 : 2 phase locking to
 $I^{(2)}(t)$ [(e), (h), and (k)],
 and
 failure of phase locking to
$I^{(3)}(t)$ [(f), (i), and (l)].
(a) Natural frequency $\omega(I)$.
 (b), (c) Sensitivity functions $\zeta(\theta,I)$ and $\xi(\theta,I)$. 
(d)--(f) Time series of the state variable $V(t)$
 of a periodically driven oscillator (red)
 and the periodic external forcing (blue).
(g)--(i)  Dynamics of the phase difference $\psi$.
  The averaged dynamics of $\psi$ is shown in the top panel, where the stable phase difference predicted by the second-order averaging of the generalized phase equation is indicated by an arrow, 
 and evolution of $\psi$
 from 20 different initial states are plotted in the bottom panel. 
 (j)--(l)  Orbits of the periodically driven oscillator (blue)
 and $I$-dependent stable limit-cycle solutions (light blue) plotted
 in three-dimensional space $(V,w,I)$.}}
 \label{fig. 2}
\end{figure*}

We set the parameters as $V_{\rm K}=-84$, $V_{\rm L}=-60$, $V_{\rm Ca}=120$, $g_{\rm K}=8$, $g_{\rm L}=2$,
$g_{\rm Ca}=4$, $C=20$, $V_1=-1.2$, $V_2=18$, $V_3=12$, $V_4=17$, and $\lambda_w=0.0667$.
For these parameters, a stable limit cycle emerges via a saddle-node on invariant circle (SNIC) bifurcation at $I \simeq 50$,
and vanishes via a Hopf bifurcation at $I \simeq 115$.
The oscillation remains generally smooth for all values of $I$.
The phase sensitivity function has the type-I shape with a positive lobe near the SNIC bifurcation,
and a sinusoidal type-II shape with both positive and negative lobes near the Hopf bifurcation~\cite{MFN}.
Thus, when the external forcing $I(t)$ is time-varying, the shape of the orbit, frequency,
and phase response properties of the oscillator can vary significantly with time.

Numerically calculated $\omega(I)$, $\zeta(\theta, I)$, and $\xi(\theta, I)$ are shown in Figs.~\ref{fig. 2} (a)--(c), and phase-locked
dynamics of the variable $V(t)$ to the periodic forcing $I(t)$ is shown in Figs.~\ref{fig. 2}(d)--(f).
Note that the oscillations are significantly deformed due to strong periodic forcing.
Figures \ref{fig. 2} (g)--(i) compare the results of the reduced phase equations with those of the direct numerical simulations.
We can confirm that the generalized phase reduction theory nicely predicts the stable phase differences $\psi$, while the conventional method does not.
The orbits of the oscillator and the cylinder $C$ of the limit cycles in three-dimensional space ($V,w,I$) are plotted in Figs. \ref{fig. 2} (j)--(l),
showing synchronous [(j) and (k)] or asynchronous (l) dynamics with the periodic forcing.

\subsection{Relaxation oscillations}

\begin{figure*}[tbp]
 \begin{center}
	\includegraphics[width=160mm]{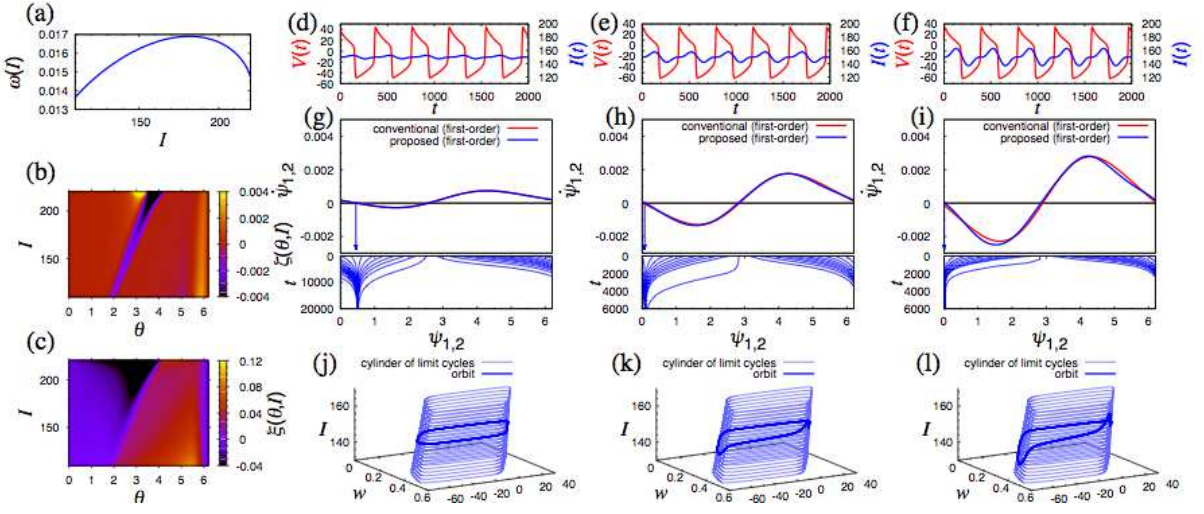}
 \end{center}
 \caption{{(Color online)
 Phase locking of the Morris-Lecar model (relaxation oscillation).
 Three types of periodically varying parameters, $I^{(j)}(t)$~:
 $q^{(j)}(\epsilon t) = 150 + \alpha^{(j)} \sin(\omega_I t) - \alpha^{(j)} \sin(2 \omega_I t)$
 and $\sigma p^{(j)}(t)=0$ with $\alpha^{(4,5,6)}=10,\,15,\,20$ and $\omega_I$ = 0.016 are used, which lead to  $1:1$ phase locking to
 $I^{(4)}(t)$ [(d), (g), and (j)], $I^{(5)}(t)$ [(e), (h), and (k)], and $I^{(6)}(t)$ [(f), (i), and (l)].
 (a) Natural frequency $\omega(I)$.
 (b), (c) Sensitivity functions $\zeta(\theta,I)$ and $\xi(\theta,I)$. 
 (d)--(f) Time series of the state variable $V(t)$
 of a periodically driven oscillator (red)
 and periodic external forcing (blue).
 (g)--(i)  Dynamics of the phase difference $\psi$
 with an arrow representing the stable phase difference (top panel)
 and evolution of $\psi$
 from 20 different initial states (bottom panel).
 (j)--(l)  Orbits of a periodically driven oscillator (blue)
 and $I$-dependent stable limit-cycle solutions (light blue) plotted
 in three-dimensional space $(V,w,I)$.}}
 \label{fig. 3}
\end{figure*}

We set the parameters as $V_{\rm K}=-84$, $V_{\rm L}=-60$, $V_{\rm Ca}=120$, $g_{\rm K}=8$, $g_{\rm L}=2$,
$g_{\rm Ca}=4.4$, $C=20$, $V_1=-1.2$, $V_2=18$, $V_3=2$, $V_4=30$, and $\lambda_w=0.004$.
For these parameters, the ML model exhibits relaxation oscillations consisting of fast and slow dynamics in an appropriate range of $I$,
and correspondingly the phase sensitivity function takes an impulse-like shape.
Numerically calculated  $\omega(I)$, $\zeta(\theta, I)$, and $\xi(\theta, I)$ are shown in Figs.~\ref{fig. 3} (a)--(c),
and the phase-locked dynamics of $V(t)$ to the periodic forcing $I(t)$ are shown in Figs.~\ref{fig. 3} (d)--(f).
Figures~\ref{fig. 3} (g)--(i) compare the results of the reduced phase equations with those of the direct numerical simulations.
The parameter $I$ was varied between $140$ and $200$.
In this case, both the conventional and generalized phase equations seem to nicely predict the stable phase difference.
As shown below, however, the conventional phase equation may actually fail to predict the oscillator dynamics in such cases.

To investigate whether the two phase equations can accurately predict dynamics of the original limit-cycle oscillator,
we further calculate the phase maps~\cite{SUCNS}, corresponding to the numerical simulations shown in Fig.~\ref{fig. 3}.
The phase map is a one-dimensional map from the phase $\theta(nT_I)$  at $t = n T_{I}$ to the phase $\theta((n+1)T_I)$
after one period of the external forcing, where $n \in \mathbb{N}$ is an integer and $T_{I}$ is the period of external forcing.
Figure~\ref{fig. 4} compares the phase maps calculated by direct numerical simulations of the original limit-cycle oscillator
with those obtained by the conventional and generalized phase equations.
These results indicate that the generalized phase equation well captures the dynamics of the oscillator, while the conventional equation does not;
it turns out that the conventional phase equation could not actually predict the oscillator dynamics in the numerical simulation of Fig.~\ref{fig. 3},
and the seemingly correct prediction of the stable phase difference was a coincidence.

\begin{figure*}[tbp]
 \begin{center}
	\subfigure[$I^{(4)}(t)$]{\includegraphics[width=80mm]{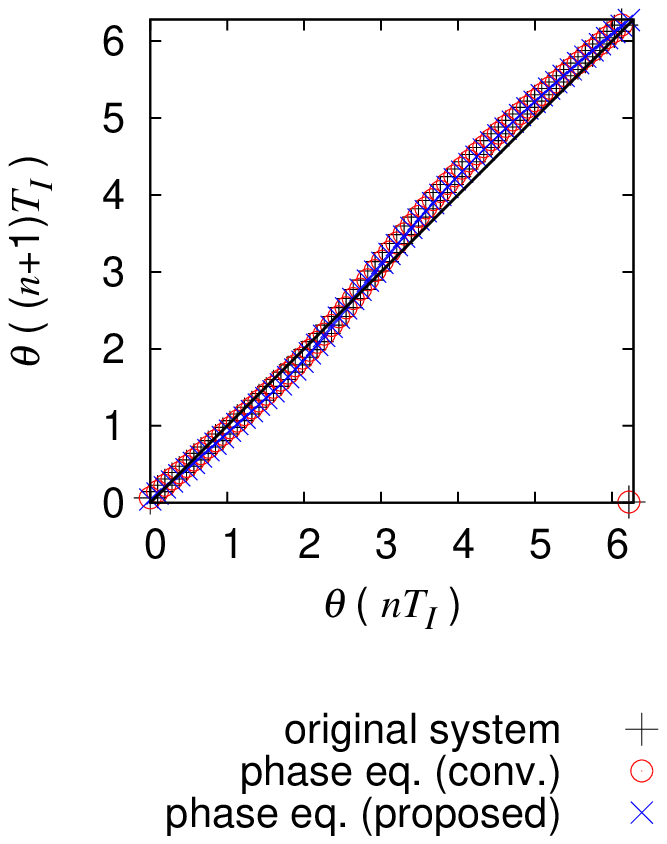}}
	\subfigure[$I^{(5)}(t)$]{\includegraphics[width=80mm]{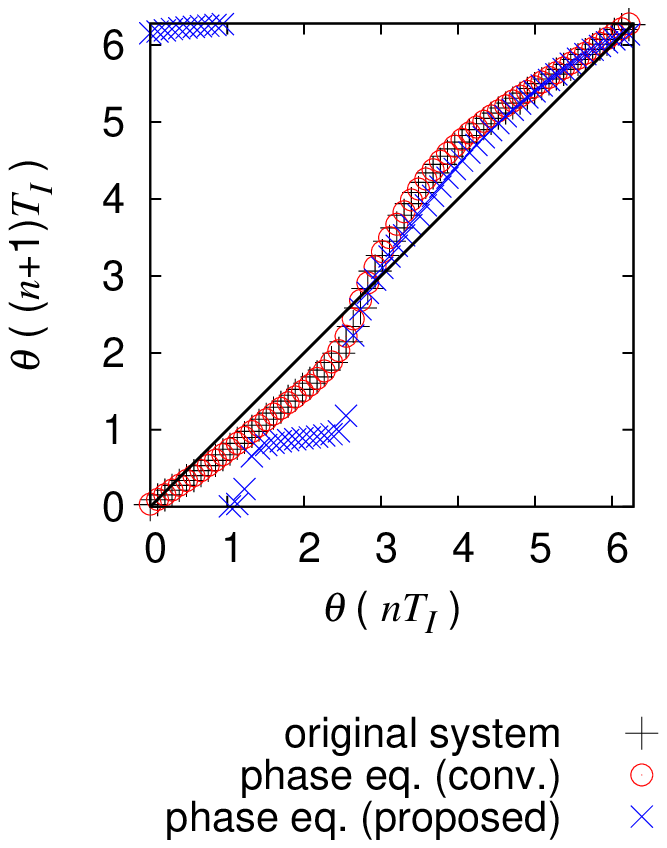}}
	\subfigure[$I^{(6)}(t)$]{\includegraphics[width=80mm]{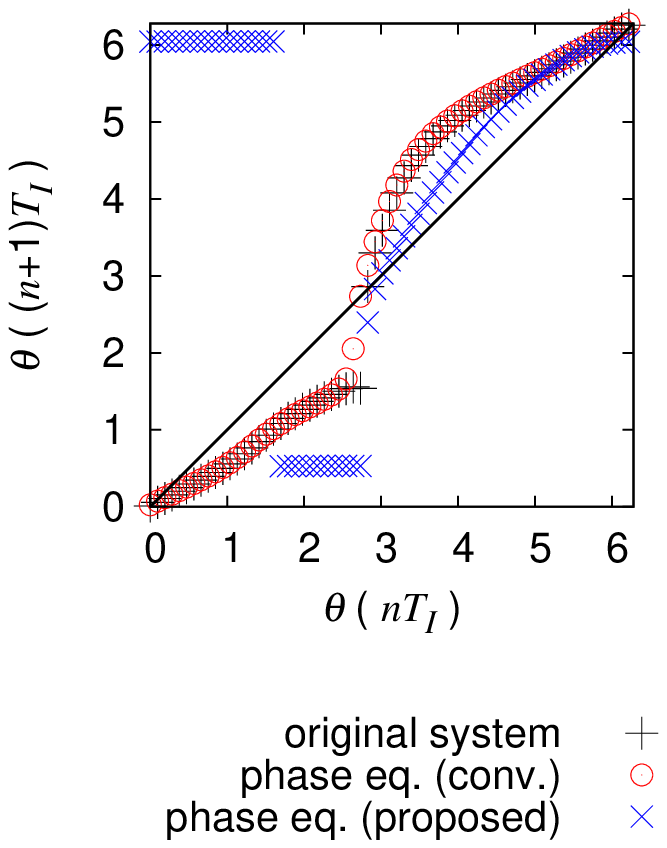}}
 \end{center}
 \caption{{(Color online) Phase maps calculated by direct numerical simulations of the original limit-cycle oscillator (black crosses)
 and by the conventional (red circles) and generalized (blue circles) phase equations.
 Results for the three types of the periodic forcing used in Fig.~\ref{fig. 3},
 i.e., (a) $I^{(4)}(t)$, (b) $I^{(5)}(t)$, and (c) $I^{(6)}(t)$, are shown.}}
 \label{fig. 4}
\end{figure*}

}


\begin{thebibliography}{99}
\bibitem{MSL}
The modified Stuart-Landau oscillator has a two-dimensional state variable $\bm{X}(t)=[x(t), y(t)]^{\top}$ and
a vector field $\bm{F}(\bm{X},\bm{I})=[e^{2I}(x-y-I)-((x-I)^2+y^2)(x-I), e^{2I}(x+y-I)-((x-I)^2+y^2)y]^{\top}$
with $\Theta(\bm{X},I)=\tan^{-1}[y/(x-I)]$, $\omega(I)=e^{2I}$,
$\bm{X}_0(\theta,I)=[I+e^{I}\cos\theta,\ e^{I}\sin\theta]^{\top}$,
$\xi(\theta,I)=e^{-I}\sin\theta$, and $\zeta(\theta,I)=2e^{2I}-e^{I}\cos\theta$.
\end{thebibliography}
\end{document}